\documentclass[acmtog,nonacm,screen]{acmart}

\author{Zihong Zhou}
\orcid{0009-0008-6052-2931}
\affiliation{%
  \institution{Dartmouth College}
  \country{USA}
}

\author{Rohan Sawhney}
\affiliation{%
  \institution{NVIDIA}
  \country{USA}
}
\orcid{0000-0002-3661-1554}

\author{Eugene d'Eon}
\affiliation{%
  \institution{NVIDIA}
  \country{Switzerland}
}
\orcid{0000-0002-3761-2989}

\author{Wojciech Jarosz}
\affiliation{%
  \institution{Dartmouth College}
  \country{USA}
}
\orcid{0000-0002-1652-0954}

\authorsaddresses{}

\usepackage{algorithm}
\usepackage{algpseudocodex}
\newcommand{\InlineIf}[2]{%
  \State \algorithmicif\ #1\ \algorithmicthen\ #2%
}

\usepackage{xcolor}
\usepackage{xfrac}
\usepackage{booktabs} 

\usepackage{rotating}
\usepackage{paralist}
\usepackage{url}
\usepackage[dvipsnames]{xcolor}
\hypersetup{
  breaklinks=true
}

\usepackage[sharp]{easylist}
\usepackage{amsmath,amssymb}

\usepackage{amsthm}
\usepackage{mathtools}
\usepackage{bm}
\usepackage{cancel}
\usepackage{wrapfig}
\usepackage{adjustbox}
\usepackage{subcaption}
\usepackage{siunitx}
\usepackage{soul}
\usepackage{array}
\usepackage{pgf}
\usepackage{tikz}
\usepackage{xspace}
\usepackage{nicefrac}
\usepackage{microtype}
\usepackage{enumitem}
\usepackage[normalem]{ulem}  
\usepackage{multirow}
\usepackage[final]{pdfcomment}  
\usepackage[outline]{contour}
\usepackage{datatool} 
\usepackage{readarray}

\usepackage[scaled=0.8]{DejaVuSansMono}
\usepackage[T1]{fontenc}

\usepackage{listings}
\usepackage[capitalize]{cleveref}

\definecolor{codegreen}{rgb}{0,0.5,0}
\definecolor{codegray}{rgb}{0.5,0.5,0.5}
\definecolor{codepurple}{rgb}{0.58,0,0.82}
\definecolor{codered}{rgb}{0.9,0,0}
\definecolor{codeblue}{rgb}{0.15,0.2,.83}

\makeatletter
\lstdefinestyle{mystyle}{
belowcaptionskip=\medskipamount, 
captionpos=t,
keepspaces=true,
numbers=left,
numbersep=1em,
tabsize=4,
showspaces=false,
showtabs=false,
breaklines=true,
frame=lines,
showstringspaces=false,
breakatwhitespace=true,
escapeinside={(*@}{@*)},
commentstyle=\color{codegreen},
keywordstyle=\color{codeblue}\bfseries,
stringstyle=\color{codered},
numberstyle=\ttfamily\fontsize{6.65}{7.5}\selectfont\color{codegray}\bfseries,
basicstyle=\ttfamily\lst@ifdisplaystyle\fontsize{6.65}{7.5}\else\fontsize{8.65}{9.5}\fi\selectfont
}

\makeatother

\newcommand{\IGNORE}[1]{}

\definecolor{DarkGreen}{rgb}{0.0,0.6,0.2}

\newcommand{\REMOVE}[1]{}

\crefname{chapter}{Chapter}{Chapters}
\crefname{section}{Sec.}{Secs.}
\crefname{subsection}{Sec.}{Secs.}
\crefname{subsubsection}{Sec.}{Secs.}
\crefname{figure}{Fig.}{Figs.}
\crefname{table}{Table}{Tables}
\crefname{listing}{Listing}{Listings}

\newcommand{\diff}{\mathrm{d}}

\newcommand{\posx}{{x}}
\newcommand{\posy}{{y}}

\newcommand{\heatadd}[1]{\textcolor{red!50!blue!75!white}{#1}}

\begin{document}

\title{Grid-Free Monte Carlo for Time-Dependent Diffusion}



\begin{abstract}
Many scientific applications require modeling how diffusive systems evolve over time, not merely their eventual steady states.
While conventional steady-state analysis of partial differential equations (PDEs) on complex geometries is already hindered by costly volumetric meshing, transient analysis further requires sequential time stepping and careful step size selection.
Grid-free Monte Carlo solvers such as walk on spheres (WoS) and walk on stars (WoSt) avoid this meshing bottleneck but remain largely limited to steady-state problems.
We generalize WoS, for pure Dirichlet problems, and WoSt, for mixed Dirichlet--Neumann problems, to heat equations with initial conditions and time-dependent source and boundary data.
We equip each random walk with a finite time budget and sample an exit time at every spatial step.
If the exit time exceeds the remaining budget, the walk samples an interior point and evaluates the initial condition; otherwise, it continues with a reduced budget, accumulating source and boundary contributions.
Our main technical contribution is a suite of kernel sampling and variance reduction techniques, including a low-bias, tabulation-free exit time sampler and efficient rejection samplers.
Unlike grid-based transient solvers, our method directly estimates the solution at any requested time without volumetric meshing or sequential time marching.
It also retains the parallel, progressive, and output-sensitive evaluation of WoS and WoSt while eliminating time step selection and temporal discretization bias entirely.
Finally, we show how sharing walks enables efficient estimates at multiple target times.

\end{abstract} 

\begin{teaserfigure}
    \centering
    \includegraphics[width=\linewidth]{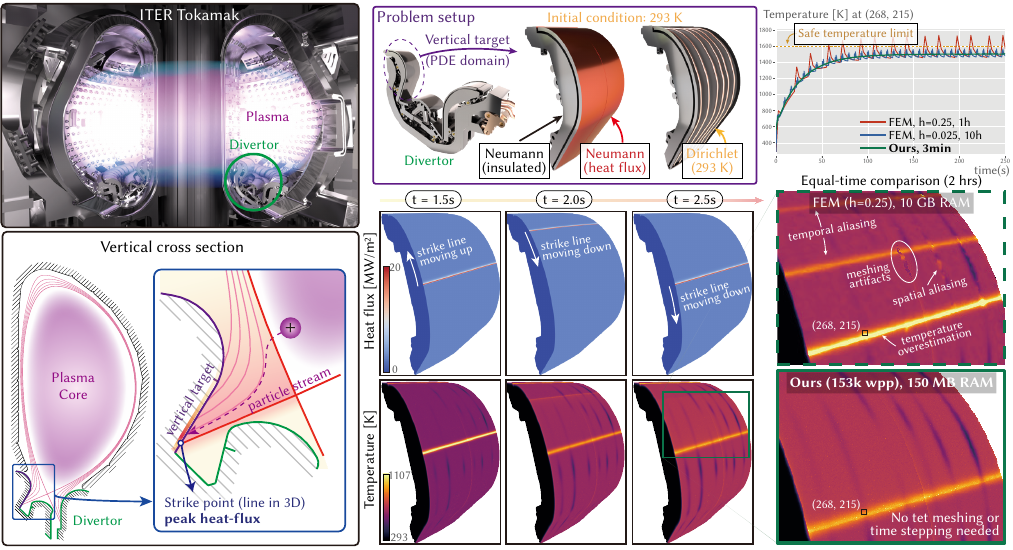}
    \caption{Our grid-free Monte Carlo solver provides robust transient thermal analysis of a vertical target in the divertor of ITER, a nuclear-fusion tokamak \citep{ITER:Divertor}. The divertor removes heat from particles escaping the confined plasma. Heat concentrates near a strike line sweeping across the target (\emph{bottom left}); we model its footprint as a spatially and temporally varying Neumann heat flux on the plasma-facing surface (\emph{center, top and middle}). Cooling pipes beneath the surface impose Dirichlet conditions, with a nonzero initial temperature throughout the target. The analysis tests whether the target remains below its safety temperature. We compare against the finite element method, whose tetrahedral volume mesh and sequential time stepping can introduce meshing artifacts and spatial and temporal aliasing (\emph{right}). These errors distort predictions and spuriously indicate a safety-limit violation. Our solver instead requires only a boundary mesh and estimates temperatures at arbitrary locations and times without time stepping. Independent random walks run in parallel and progressively refine estimates; computation can focus on selected probes, with each walk estimating temperatures at all requested times (\emph{top right}).}
    \label{fig:teaser}
\end{teaserfigure}

%
%

\ccsdesc[500]{Mathematics of computing~Partial differential equations}
\ccsdesc[500]{Mathematics of computing~Integral equations}
\ccsdesc[500]{Mathematics of computing~Probabilistic algorithms}

\keywords{Numerical Methods, Monte Carlo Methods, Partial Differential Equations, Walk on Spheres, Functional Expansion}
\maketitle

\section{Introduction}
\label{sec:intro}
Many physical systems are governed by diffusion: heat spreads through solids, electromagnetic fields penetrate and attenuate within conductive materials, and chemical species disperse through media.
In practice, their evolution toward equilibrium is often as important as the equilibrium itself.
For instance, in thermal design, resolving transient behavior reveals when and where heat accumulates, helping engineers prevent overheating in systems ranging from consumer electronics \citep{Zhan:2007:ThermallyAwareDesign} and space-based data centers \citep{Celeroton:2025:Space} to plasma-facing divertor components in fusion reactors (\cref{fig:teaser}).
The heat equation, known more generally as the diffusion equation, provides a canonical model for this transient evolution \citep{Carslaw:1959:Conduction}.
Its solution is determined by an initial condition specifying the starting distribution, a source term describing generation or removal within the domain, and boundary conditions prescribing the field's value or flux at the boundary.

In this work, we seek to solve the following mixed initial-boundary value problem (IBVP) for the inhomogeneous heat equation:
\begingroup
\setlength{\fboxsep}{6pt}
\begin{equation}\label{eq:goalPDE}
\boxed{
\begin{alignedat}{3}
\frac{\partial u}{\partial t}(\posx,t)
&= \kappa\Delta u(\posx,t) + f(\posx,t),
&\qquad & \posx \in \Omega,
&\quad & t>0, \\
u(\posx,0)
&= u_0(\posx),
& & \posx \in \Omega,
& & t=0, \\
u(\posx,t)
&= g(\posx,t),
& & \posx \in \partial\Omega_{\mathrm{D}},
& & t>0, \\
\frac{\partial u}{\partial n_{\posx}}(\posx,t)
&= h(\posx,t),
& & \posx \in \partial\Omega_{\mathrm{N}},
& & t>0.
\end{alignedat}
}
\end{equation}
\endgroup
Here, $\Omega\subset\mathbb{R}^d$, with $d\in\{2,3\}$, is the domain, and $u(\posx,t)$ denotes an evolving scalar field at $\posx\in\Omega$ and time $t\geq0$.
The constant $\kappa>0$ is the diffusivity, and $\Delta$ the spatial Laplacian.
The boundary $\partial\Omega$ is partitioned into Dirichlet and Neumann portions, $\partial\Omega_{\mathrm D}$ and $\partial\Omega_{\mathrm N}$, on which $g$ prescribes the field value and $h$ its outward normal derivative, respectively.
The function $u_0$ specifies the initial condition at $t=0$ and $f$ the interior source; $f$, $g$, and $h$ may vary with time.
When the prescribed inputs admit a steady state solution, $u(\posx,t)$ approaches it as $t\to\infty$ (\cref{fig:comparison-steady-state-transient-dirichlet,fig:comparison-steady-state-transient-mixed}).

Finite element methods (FEM) are widely used for IBVPs such as \cref{eq:goalPDE}.
Yet, even for steady-state problems, applying FEM to complex domains requires a volumetric mesh that resolves geometric detail, which can be expensive, memory intensive, and failure-prone \citep{Sawhney:2020:Monte,Sawhney:2023:Walk,Miller:2024:Walkin}.
Transient problems compound this difficulty: the mesh must also resolve localized interior features induced by initial conditions and time-dependent sources.
Under-resolving these features produces spatial aliasing, analogous to under-resolving variation induced by variable coefficients \citep{Sawhney:2022:Gridfree}.
Beyond these spatial demands, FEM converts the heat equation into a large system of coupled ordinary differential equations (ODEs), advanced through sequential time stepping.
Selecting an integrator and step size requires balancing accuracy, stability, and cost, but even a carefully chosen scheme introduces temporal discretization error (see \cref{sec:results:fem}).

The burden of spatial discretization motivated renewed interest in the classical walk on spheres (WoS) method, a grid-free Monte Carlo solver for Laplace and Poisson problems with Dirichlet boundary conditions \citep{Muller:1956:Continuous,Sawhney:2020:Monte}.
Walk on stars (WoSt) subsequently extended this framework to mixed Dirichlet--Neumann problems \citep{Sawhney:2023:Walk}.
Both methods recursively express the solution at a query point in terms of values at random exit locations on a local sphere or star-shaped region.
At each step, distance queries against the boundary determine the region size, and repeated exit sampling produces a random walk whose expected contribution yields the solution.
This local, query-driven construction avoids volumetric meshing and global linear solves, while independent samples enable parallel, progressive, and output-sensitive evaluation.
Yet despite numerous extensions (reviewed in \cref{sec:related-grid-free-monte-carlo}), these solvers remain largely limited to steady-state PDEs.

\begin{figure}[t]
    \centering
    \includegraphics[width=\linewidth]{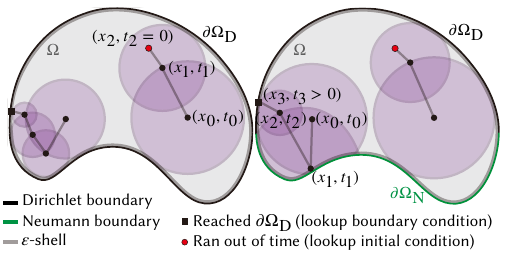}
    \caption{To solve heat equations with pure Dirichlet (\emph{left}) or mixed Dirichlet--Neumann boundary conditions (\emph{right}), we modify the walk on spheres and stars algorithms (resp.) to sample an exit time with each spatial step of a random walk. Walks are terminated either when they reach the epsilon shell around the Dirichlet boundary, or when they run out of time inside the domain where we evaluate the initial condition.
    }
    \label{fig:concept-random-walk}
\end{figure}

To overcome this limitation, we generalize WoS and WoSt to solve the IBVP in \cref{eq:goalPDE} without \emph{any} temporal discretization.
We augment each random walk with a time budget and sample an exit time alongside the exit location at every sphere or star step (\cref{fig:concept-random-walk}).
If the exit time fits the budget, the walk advances with a reduced budget; otherwise, it terminates at an interior point sampled from the time-conditioned heat kernel and evaluates the initial condition (Secs.~\ref{sec:method:initial}--\ref{sec:method:sampling}).
A walk reaching the Dirichlet boundary instead evaluates the boundary data, while source and Neumann contributions are accumulated along the walk by sampling their associated space-time kernels (Secs.~\ref{sec:method:wost}--\ref{sec:method:inhomogeneous}).
Crucially, this temporal machinery requires no changes to the distance, silhouette, or ray-intersection queries used by existing WoS and WoSt implementations.

Although finite-time random walks date to \citet{HajiSheikh:1966:Floating}, their exit time sampler relies on tabulation and CDF inversion.
Their treatment of general initial conditions, sources, and Neumann data is likewise limited.
To address these shortcomings, we develop a tabulation-free exit time sampler with very low bias and efficient rejection samplers for temporal and radial variables associated with prescribed inputs (\cref{sec:method}).
These rejection samplers exploit distinct short-, intermediate-, and long-time behavior of the heat kernel and its derivatives for robustness across time scales (\cref{fig:exit-time-heat-kernel-sampling}).
We also develop variance reduction strategies and amortize geometric queries across multiple times using shared walks (\cref{sec:efficiency}).

\begin{figure}[t]
    \centering
    \includegraphics[width=\linewidth]{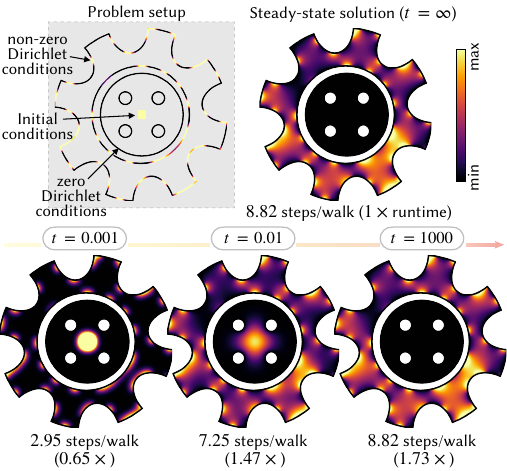}
    \caption{Given a fixed time budget $t$, the solution to a heat equation interpolates between initial condition values prescribed at $t = 0$ (\emph{top left}) and diffused boundary values at $t = \infty$ (\emph{top right}). Our method computes the solution for any $t$ (\emph{bottom row}), taking fewer steps per walk for finite times compared to standard walk on spheres, as walks may terminate inside the domain. Consequently, method efficiency is time-dependent, determined by both the average walk length and the overhead of exit time sampling.}
    \label{fig:comparison-steady-state-transient-dirichlet}
\end{figure}

Together, these developments extend WoS and WoSt to time-dependent problems without volumetric or temporal discretization, avoiding aliasing from insufficient mesh resolution or large time steps (\cref{fig:teaser}).
Monte Carlo noise instead decreases predictably with more walks.
At short times, walks require fewer steps than their steady-state counterparts because they often exhaust their time budgets before reaching the boundary (\cref{fig:comparison-steady-state-transient-dirichlet,fig:comparison-steady-state-transient-mixed}).
At long times, they approach steady-state behavior with modest additional cost for time-dependent kernel sampling.
Shared walks also improve efficiency when evaluating multiple target times.

Our method is particularly attractive for sparse or localized queries, short time horizons, and progressive evaluation.
However, like other Monte Carlo methods, it produces noisy estimates and is therefore less competitive when dense global solutions are required, a regime in which conventional grid-based solvers remain highly effective.
Extending recent sample-caching techniques for steady-state Monte Carlo solvers \citep{Qi:2022:Bidirectional,Miller:2023:Boundary,Bakbouk:2023:Mean,Zhou:2025:Harmonic} to the time-dependent setting offers a promising direction for improving dense transient solution estimates.

\begin{figure}[t]
    \centering
    \includegraphics[width=\linewidth]{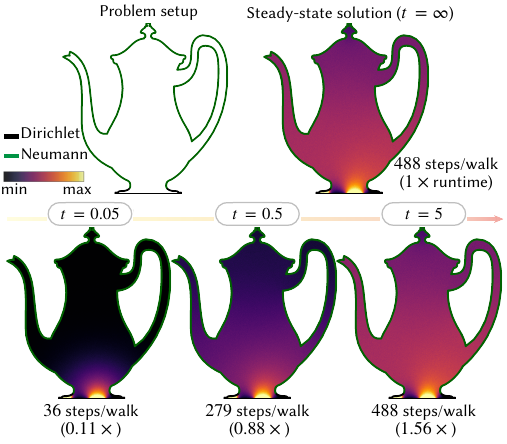}
    \caption{
    An analogous experiment to \cref{fig:comparison-steady-state-transient-dirichlet} using walk on stars with time-varying Dirichlet and Neumann boundary conditions, and zero initial conditions. As before, our finite-time algorithm requires fewer steps per walk than the steady-state estimator, with overall efficiency determined by both average walk length and exit time sampling overhead.}
    \label{fig:comparison-steady-state-transient-mixed}
\end{figure}

\section{Related Work}
\label{sec:related}
We review related PDE solvers in two broad categories: grid-based deterministic methods and grid-free Monte Carlo methods, with particular emphasis on approaches for time-dependent problems.

\subsection{Grid-Based Deterministic Solvers}
\label{sec:related-grid-based-deterministic}

Conventional deterministic solvers typically discretize space.
Finite difference methods (FDM) approximate differential operators on a regular grid \citep{LeVeque:2007:FiniteDifference}, while finite element methods (FEM) approximate the solution in a finite-dimensional basis over a volumetric mesh \citep{Zienkiewicz:1977:finite}.
For FEM, robustly generating a volumetric mesh that resolves complex geometry can itself be expensive and failure-prone \citep{Si:2015:TetGen,Hu:2018:TetWild,Hu:2020:fTetWild}.
Even once constructed, the mesh must resolve spatial variation induced by initial and boundary conditions and sources.
Under-resolving causes spatial aliasing, yet sufficient resolution is rarely known a priori and may require costly, memory-intensive refinement.
Alternatively, boundary element methods (BEM) discretize only the domain boundary but produce dense linear systems whose memory and factorization costs grow rapidly with boundary resolution.
Nonconstant volumetric data, including initial conditions and sources, erode this boundary-only advantage by introducing domain integrals requiring volume quadrature or auxiliary approximations \citep{Zienkiewicz:1977:Coupling,Costabel:1987:symmetric,Ingber:1992:Parabolic,Partridge:2012:Dual}.

\paragraph{Time-Dependent Problems}
After spatial discretization, the heat equation becomes a coupled ODE system typically advanced by time stepping.
Explicit schemes such as Forward Euler and fourth-order Runge--Kutta are conditionally stable, with step size constrained by the finest spatial resolution, while implicit schemes such as Backward Euler and Crank--Nicolson allow larger steps but require a global solve at each step \citep{Wanner:1996:Solving,Thomee:2007:Galerkin}.
Choosing an integrator and step size therefore involves a problem-dependent trade-off: large steps may under-resolve transient dynamics and time-dependent data, causing temporal aliasing, whereas small steps reduce error at higher cost.
Because FEM supports the initial condition, source term, and mixed boundary conditions in \cref{eq:goalPDE}, it serves as our principal deterministic baseline (\cref{sec:results:fem}).

Two BEM formulations offer alternatives to ODE integration---time-domain boundary integral equations and Laplace-domain methods.
In both cases, retaining a boundary-only formulation generally requires zero or constant initial conditions and sources; general volumetric inputs introduce domain integrals.
Time-domain BEM expresses boundary operators as history convolutions.
Without specialized acceleration, storing and evaluating this history grows increasingly expensive with the number of time steps \citep{Wrobel:1979:Boundary,Lubich:1992:Time,Lubich:2004:Convolution}.
The Laplace transform instead converts the transient heat equation into independent modified Helmholtz problems at selected transform parameters, which can be solved in parallel and numerically inverted at multiple times without sequential time marching \citep{Rizzo1970:Method}.
However, accurate inversion may require many carefully placed samples across relevant time scales.
More critically, numerical inversion is notoriously ill-conditioned and susceptible to numerical cancellation \citep{Epstein:2008:Bad,Kuhlman:2013:InverseLaplace}.

In contrast, our time-dependent WoS and WoSt extensions evaluate solutions in continuous spacetime without volumetric meshing, time stepping, or Laplace inversion.
They are output-sensitive in space and time, supporting individual spacetime queries or simultaneous evaluation at multiple target times without sequential evolution from $t=0$ (\cref{sec:efficiency:shared}).

\subsection{Grid-Free Monte Carlo Solvers}
\label{sec:related-grid-free-monte-carlo}

Stochastic representations such as the Feynman--Kac formula express PDE solutions as expectations over Brownian paths \citep{Oksendal:2003:Stochastic}.
Direct Monte Carlo implementations discretize these paths into small time steps \citep{Higham:2001:Algorithmic}.
In complex domains, this creates a trade-off: small steps resolve boundary interactions but produce long walks, whereas large steps may miss crossings and introduce geometric bias, as demonstrated in \citet[Section 7.1.3]{Sawhney:2022:Gridfree} and \citet[Section 6.3]{Sawhney:2023:Walk}.

Grid-free random walks avoid this trade-off by sampling first exits from simple regions.
For steady-state Laplace equations $\Delta u=0$, walk on spheres recursively jumps between spheres contained in the domain, sampling exit locations uniformly through the mean-value property of harmonic functions \citep{Muller:1956:Continuous}.
WoS requires neither a volumetric mesh nor discretization of Brownian paths, introducing a controllable bias when walks terminate within an $\varepsilon$-shell of the boundary.
This bias can be reduced using debiasing or conformal techniques \citep{Misso:2022:Unbiased,Paul:2025:Conformal}.

Since its introduction to computer graphics by \citet{Sawhney:2020:Monte}, WoS has been extended to a broader set of boundary conditions through walk on stars \citep{Sawhney:2023:Walk,Miller:2024:Walkin,Bao:2026:Monte}, spatially varying coefficients \citep{Sawhney:2022:Gridfree}, participating media \citep{Miller:2025:Participating}, and robust gradient estimation \citep{Yu:2025:Robust}.
Recent work has also developed variance reduction techniques \citep{Nabizadeh:2021:Kelvin,Qi:2022:Bidirectional,Bakbouk:2023:Mean,Miller:2023:Boundary,Li:2023:Neural,Li:2024:Neural,Huang:2025:Guiding,Zhou:2025:Harmonic,Bao:2025:OffCentered}, differentiable solvers for inverse problems \citep{Miller:2024:Differential,Yu:2024:Differential,Yilmazer:2024:Solving,Himmler:2026:Biharmonic}, and applications in fluid dynamics, thermal imaging, and rendering \citep{Rioux-Lavoie:2022:Monte,Jain:2024:Neural,Sugimoto:2024:Velocitybased,Bati:2023:Coupling,Wu:2025:Unbiased}.
See \citet{Sawhney:2025:State,Sawhney:2025:StateURL} for a comprehensive overview.
These developments, however, focus primarily on steady-state problems.

\paragraph{Time-Dependent Problems}
Early grid-free Monte Carlo solvers for the heat equation were introduced by \citet{HajiSheikh:1966:Floating,HajiSheikh:1967:Solution} as ``floating random walks.''
Their central contribution was a sampler for the first exit time from a sphere.
By rescaling the exit time with the sphere radius, they reduced sampling to numerical inversion of a single one-dimensional cumulative distribution function (CDF).
Constructing this CDF accurately, however, can be numerically unstable at short times.
Their treatment of other prescribed inputs in \cref{eq:goalPDE} is also less general: nonconstant initial conditions require an under-specified tabulation procedure, sources are restricted to constants, and Neumann conditions are approximated using finite differences---an approach prone to high bias and slow runtimes on complex domains \citep[Sec.~6.3]{Sawhney:2023:Walk}.

Subsequent work explored tabulation-free alternatives.
\citet{Deaconu:2013:Hitting,Deaconu:2017:Walk} derived the walk on moving spheres (WoMS) algorithm from Bessel-process hitting times, enabling exit time sampling without precomputed tables.
\citet{Sabelfeld:2017:Random} briefly suggests applying alternating-series rejection \citep{Devroye:1986:NonUniform} to the ball exit time distribution but omits the required normalization constant and enough detail to assess the sampler.
Rather than sample the exit time from a fixed sphere, the walk on heat balls (WoHB) method of \citet{Deaconu:2018:Initial} couples each temporal step to the radius of a spacetime heat ball.
This construction takes smaller spatial steps than a maximal WoS sphere and cannot terminate exactly at the initial time; its implementation instead uses a temporal $\varepsilon$-shell, introducing additional bias.
Consequently, WoHB requires more steps and exhibits greater bias than our method (\cref{sec:results:wohb}).
Concurrent work extends this heat ball construction to mixed Dirichlet--Neumann conditions and source terms through walk on heat stars (WoHSt) \citep{Bao:2026:HeatStars}.
Building directly on WoHB, WoHSt likewise uses non-maximal star-shaped regions and a temporal $\varepsilon$-shell for termination.

\citet{DeLambilly:2023:Heat} take a different grid-free approach, approximating transient diffusion by exponentially interpolating between the initial condition and a steady-state WoS solution using a decay rate derived from the domain's bounding box.
This approximation is least accurate at short times, when heat is localized rather than uniformly approaching steady state.
It also does not support source or Neumann data.
Finally, walk on boundary (WoB) solves heat equations with nonconstant initial conditions using ray tracing to construct walks between points on the domain boundary \citep{Sabelfeld:1994:Random,Sugimoto:2023:Practical,Sugimoto:2024:Velocitybased}.
On non-convex domains, however, selecting among multiple ray intersections can produce large path weights and unbounded variance, while the sign-changing heat flux kernel causes severe cancellation.
Despite tracing substantially more walks per unit time, WoB remains extremely noisy at longer times in our experiments (\cref{sec:results:wob}).

Against this backdrop, we develop robust evaluation and sampling methods for the heat kernel and exit time distribution in two and three dimensions (\cref{sec:method}).
Their spectral representations converge rapidly at long times but require many terms and suffer floating-point cancellation at short times \citep{Carslaw:1959:Conduction,Grebenkov:2013:Efficient,Rupprecht:2015:Exit,Malecki:2016:FourierBessel,Serafin:2017:Exit}.
We therefore switch between compact spectral expansions at long times and method-of-images approximations at short times.
Since the heat kernel governs exit times, initial conditions, source contributions, and Neumann data, directly or through integration or differentiation, its robust evaluation provides a foundation for sampling.
For exit time sampling, we replace tabulation with a finite sum of exactly sampled exponential variables and a moment-matched Gamma approximation of the tail, yielding very low bias at substantially lower cost than WoMS (\cref{fig:exit-time-heat-kernel-sampling}).
We also construct efficient, numerically accurate rejection samplers for temporal and radial variables associated with the prescribed inputs in \cref{eq:goalPDE}.
Finally, we develop targeted sampling strategies for localized initial conditions and time-independent source and Neumann data (\cref{sec:efficiency}).

\section{Background}
\label{sec:background}

After establishing notation, we review the grid-free Monte Carlo methods underlying our work: walk on spheres and walk on stars for steady-state Dirichlet and mixed Dirichlet--Neumann problems, respectively, followed by floating random walks for the heat equation \citep{HajiSheikh:1966:Floating}.

\subsection{Notation}
\label{sec:bg:notation}

\paragraph{Spatial notation}
For a spatial region $A\subset\mathbb{R}^d$, $|A|$ and $|\partial A|$ denote its $d$-dimensional volume and $(d-1)$-dimensional boundary measure, respectively.
We use $x$ for a query point or current walk position, $y$ for an interior point, and $z$ for a boundary point; each lies in $\mathbb{R}^d$.
We write $B(x,R)$ for the ball of radius $R$ centered at $x$ and $\partial B(x,R)$ for its boundary sphere.
For $y\in B(x,R)$, we define the radial distance $r=\lVert x-y\rVert$ and normalized radius $\rho=r/R\in[0,1]$.
The $\varepsilon$-shell inside the domain $\Omega$ is $\{x\in\Omega:\operatorname{dist}(x,\partial\Omega)<\varepsilon\}$.
Finally, $\mathbb{P}$ denotes probability, and $p^A(x)$ is a probability density function (PDF) on $A$ relative to the appropriate volume or boundary measure.

\paragraph{Temporal notation}
The variable $t\geq0$ denotes the time at which the PDE solution is evaluated.
A random walk begins with the time budget $t_0=t$, with $t_k$ denoting its remaining budget at step $k$.
We use $\eta$ for elapsed time, either as an integration variable or as a random variable, and $\eta_k$ for the elapsed time sampled at step $k$.
For a ball of radius $R$, we define the dimensionless elapsed time $\tau=\kappa\eta/R^2$, and use it to distinguish between short-, intermediate-, and long-time regimes during kernel evaluation and sampling (\cref{sec:method:sampling}).

\paragraph{Units}
Except in \cref{fig:teaser}, we normalize coordinates so each domain lies within $[0,1]^d$.
To interpret normalized results in physical units, let one normalized unit correspond to a physical length $L$.
Since $x_{\mathrm{phys}}=Lx$, the Laplacian transforms as $\Delta_{\mathrm{phys}}=L^{-2}\Delta$, and the heat equation relates physical and normalized time by $t_{\mathrm{phys}}=tL^2/\kappa$.
For $L=1\,\mathrm{m}$ and $\kappa=10^{-4}\,\mathrm{m}^2/\mathrm{s}$, representative of highly conductive metals such as copper and aluminum, $t=10^{-3}$ corresponds to $t_{\mathrm{phys}}=10\,\mathrm{s}$ (\cref{fig:comparison-steady-state-transient-dirichlet}).
We generally assume unit diffusivity, $\kappa=1$; \cref{sec:method:diffusivity} explains how to handle other positive constant diffusivities.

\subsection{Walk on Spheres}
\label{sec:bg:wos}

With time-independent data and unit diffusivity, the steady-state, pure Dirichlet specialization of \cref{eq:goalPDE} is the Poisson equation $-\Delta u=f$ in $\Omega$, with $u=g$ on $\partial\Omega$.
For any ball $B(x,R)\subset\Omega$, Green's representation formula gives \citep{Muller:1956:Continuous,Sawhney:2020:Monte}
\begin{align}\label{eq:meanValueIntegral}
    u(x) = \int_{\partial B(x,R)} P^B(x,z)u(z)\diff z + \int_{B(x,R)} G^B(x,y)f(y)\diff y,
\end{align}
where $G^B$ is the harmonic Green's function of the ball and $P^B(x,z):=-\partial G^B(x,z)/\partial n_z$ is its Poisson kernel.
Because the ball is centered at $x$, the Poisson kernel is uniform, with $P^B(x,z)=1/|\partial B(x,R)|$.
Explicit formulas for both kernels in two and three dimensions are given by \citet[Appendix~A]{Sawhney:2023:Walk}.
When $f=0$, the volume term vanishes and \cref{eq:meanValueIntegral} reduces to the mean-value property for harmonic functions.

At step $k$ of a random walk with current position $x_k$, Monte Carlo integration of \cref{eq:meanValueIntegral} yields the one-sample recursive estimator
\begin{align}\label{eq:WoSEstimator}
    \widehat{u}(x_k) = \frac{P^B(x_k, x_{k+1}) \widehat{u}(x_{k+1})}{p^{\partial B}(x_{k+1})} + \frac{G^B(x_k,y_k) f(y_k)}{p^{B}(y_k)}.
\end{align}
Here, $x_{k+1}$ and $y_k$ are sampled from densities $p^{\partial B}$ on $\partial B(x_k,R_k)$ and $p^B$ in $B(x_k,R_k)$, respectively.
WoS chooses the largest ball centered at $x_k$ and samples $x_{k+1}$ uniformly, so $p^{\partial B}(x_{k+1})=P^B(x_k,x_{k+1})$ and the boundary weight cancels.
Recursively applying the estimator produces a random walk through the domain, while the volume samples accumulate source contributions, for example by importance sampling the Green's function \citep[Sec.~4.2]{Sawhney:2020:Monte}.
The walk terminates upon entering the $\varepsilon$-shell, where it evaluates $g$ at the closest point on the Dirichlet boundary.

\subsection{Walk on Stars}
\label{sec:bg:wost}

Walk on stars extends WoS to the steady-state specialization of \cref{eq:goalPDE} with mixed Dirichlet--Neumann boundary conditions \citep{Sawhney:2023:Walk}.
WoSt constructs a star-shaped region $\mathrm{St}(x,R)=\Omega\cap B(x,R)$ centered at the current walk position $x$, whose boundary comprises a spherical portion $\partial\mathrm{St}_{\mathrm B}(x,R)$ and a portion $\partial\mathrm{St}_{\mathrm N}(x,R)\subseteq\partial\Omega_{\mathrm N}$ of the Neumann boundary.
Its radius $R$ is the minimum of the distance to the Dirichlet boundary and the distance to the closest silhouette point on $\partial \Omega_{\mathrm N}$.
This choice ensures that $\mathrm{St}(x,R)$ is star-shaped with respect to $x$; see \citet[Sec.~5]{Sawhney:2023:Walk} for algorithms that compute these distances on mesh-based geometry.
Specializing Green's representation formula to this region and replacing the normal derivative on $\partial\mathrm{St}_{\mathrm N}(x,R)$ with the prescribed Neumann data $h$ then gives
\begin{align}\label{eq:wostIntegral}
    \alpha(x)u(x) &= \int_{\partial\mathrm{St}(x,R)} P^B(x,z)u(z)\diff z + \int_{\mathrm{St}(x,R)} G^B(x,y)f(y)\diff y\notag\\
                  &\quad+ \int_{\partial\mathrm{St}_{\mathrm N}(x,R)} G^B(x,z)h(z)\diff z,
\end{align}
where $\alpha(x)=1$ for an interior point and $\alpha(x)=1/2$ at a smooth Neumann boundary.
No normal derivative contribution remains on $\partial\mathrm{St}_{\mathrm B}$, since $G^B=0$ on $\partial B(x,R)$.

As in WoS, recursively sampling the first integral generates a random walk, while the remaining integrals accumulate source and Neumann contributions.
Because $\mathrm{St}(x,R)$ is star-shaped with respect to $x$, each sampled direction identifies a unique first intersection with its boundary.
In particular, uniform direction sampling importance samples the Poisson kernel without requiring explicit surface-area sampling \citep{Veach:1995:Optimally}.
A sampled ray either reaches the spherical portion $\partial\mathrm{St}_{\mathrm B}$ and advances the walk through the domain, or reaches the Neumann portion $\partial\mathrm{St}_{\mathrm N}$ and continues from that boundary point.
The walk terminates upon entering the Dirichlet $\varepsilon$-shell, as in WoS.

\subsection{Floating Random Walks}
\label{sec:bg:floating}

\citet{HajiSheikh:1966:Floating,HajiSheikh:1967:Solution} extended WoS to the homogeneous heat equation ($f=0$) with Dirichlet boundary conditions under the name floating random walks.
The term ``floating'' distinguishes these continuous-space walks from walks tied to a fixed grid.
For a constant initial condition $u_0$ and unit diffusivity, the solution over a ball $B(x,R)\subset\Omega$ satisfies
\begin{align}\label{eq:floatingIntegralConstantIC}
    u(x,t) &= \int_0^t\int_{\partial B(x,R)} P^B(x,z,\eta)u(z,t-\eta)\diff z\diff\eta\notag\\
           &+ u_0\int_{B(x,R)}G^B(x,y,t)\diff y,
\end{align}
where $G^B(x,y,t)$ is the heat kernel of the ball and $P^B(x,z,\eta)$ is its time-dependent Poisson kernel.
Since $u_0$ is constant, it factors out of the spatial integral of the heat kernel.

\begin{figure}[t]
    \centering
    \includegraphics[width=\linewidth]{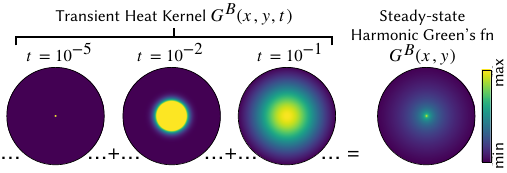}
    \caption{We use the heat kernel $G^B(x, y, t)$ and its derivatives to sample exit times and locations over a ball $B$. For a given time $t$, this kernel describes how heat spreads from a single Dirac impulse at point $x$ to any other location $y$ inside $B$. Standard walk on spheres and walk on stars instead use the harmonic Green's function $G^B(x, y)$, which equals the time-integrated value of the heat kernel from $t = 0$ to $t = \infty$.}
    \label{fig:concept-heat-kernel}
\end{figure}

In three dimensions, the heat kernel centered at $x$ is
\begin{align}\label{eq:HeatKernel3D}
    G^B(x,y,t) &= \frac{1}{2R^2r} \sum_{n=1}^{\infty} n\sin\left(\frac{n\pi r}{R}\right) \exp\left(-\frac{n^2\pi^2t}{R^2}\right),
\end{align}
where $r=\lVert x-y\rVert$. It is related to the harmonic Green's function in \cref{eq:meanValueIntegral} by
\begin{align}\label{eq:heatKernelToHarmonicGreen}
    G^B(x,y) &= \int_0^\infty G^B(x,y,\eta)\diff\eta.
\end{align}
Thus, the harmonic Green's function accumulates the heat kernel over all elapsed times (\cref{fig:concept-heat-kernel}).

As in the steady-state case, the time-dependent Poisson kernel is the negative outward normal derivative of the heat kernel:
\begin{align}\label{eq:transientPoissonKernel}
    P^B(x,z,\eta) &:= -\frac{\partial G^B}{\partial n_z}(x,z,\eta).
\end{align}
For a ball centered at $x$, it factorizes into independent exit location and exit time distributions:
\begin{align}\label{eq:PoissonKernelSeparation}
    P^B(x,z,\eta) &= \frac{1}{|\partial B(x,R)|}H(\eta;R),\notag\\
    H(\eta;R) &= \frac{2\pi^2}{R^2}\sum_{n=1}^{\infty}(-1)^{n-1}n^2\exp\left(-\frac{n^2\pi^2\eta}{R^2}\right).
\end{align}
Here, $H(\eta;R)$ is the exit time density in three dimensions and the exit location is uniform on the sphere.
The spectral series for both $G^B$ and $H$ converge rapidly at long times but require increasingly many terms for accurate evaluation as $t/R^2$ or $\eta/R^2$ approaches zero.
We provide the corresponding two-dimensional heat kernel and exit time density formulas in the supplemental document.

Spatially, floating random walks proceed exactly as WoS.
At step $k$, the algorithm constructs the largest ball $B(x_k,R_k)$ and samples an exit time $\eta_k$ from a tabulated representation of $H(\cdot;R_k)$.
If $\eta_k\leq t_k$, it samples $x_{k+1}$ uniformly on the sphere, updates the remaining budget to $t_{k+1}=t_k-\eta_k$, and continues.
Otherwise, the walk runs out of time and returns the constant initial value $u_0$.
As in WoS, a walk that first enters the Dirichlet $\varepsilon$-shell instead returns the prescribed boundary value at its current remaining time.

\section{Method}
\label{sec:method}

Building on floating random walks, we develop our solver for the IBVP in \cref{eq:goalPDE} incrementally, beginning with time-dependent WoS for pure Dirichlet problems with general initial conditions (\cref{sec:method:initial}).
We then present our tabulation-free exit time and time-conditioned heat kernel samplers required to perform these random walks efficiently (\cref{sec:method:sampling}).
Next, we extend the formulation to mixed Dirichlet--Neumann boundary conditions using WoSt (\cref{sec:method:wost}), incorporate time-dependent source and Neumann data (\cref{sec:method:inhomogeneous}), and conclude by treating arbitrary positive constant diffusivities (\cref{sec:method:diffusivity}).

\subsection{Walk on Spheres with General Initial Conditions}
\label{sec:method:initial}

We first consider the homogeneous, pure Dirichlet specialization of \cref{eq:goalPDE} with unit diffusivity and a nonconstant initial condition $u_0(x)$.
Unlike in the constant case of \cref{eq:floatingIntegralConstantIC}, $u_0$ cannot be factored out of the heat kernel integral, which becomes
\begin{align}\label{eq:initialConditionIntegral}
    \int_{B(x,R)}G^B(x,y,t)u_0(y)\diff y.
\end{align}
Letting $(x_k,t_k)$ denote a random walk's current position and remaining time at step $k$, Monte Carlo integration then yields the one-sample recursive estimator
\begin{align}\label{eq:generalDirichletEstimator}
    \widehat{u}(x_k,t_k) &=
    \frac{P^B(x_k,x_{k+1},\eta_k)\widehat{u}(x_{k+1},t_k-\eta_k)}{\mathbb{P}^{\partial B}p^{\partial B}(x_{k+1},\eta_k)}\notag\\
    &+ \frac{G^B(x_k,y_{k+1},t_k)u_0(y_{k+1})}{\mathbb{P}^{B}p^{B}(y_{k+1})}.
\end{align}
Here, $\mathbb{P}^{\partial B}$ and $\mathbb{P}^{B}$ denote the probabilities of evaluating the boundary and initial condition terms, respectively.
When the boundary term is evaluated, $(x_{k+1},\eta_k)$ is drawn from the joint density $p^{\partial B}$ on $\partial B(x_k,R_k)\times[0,t_k]$;
when the initial condition term is evaluated, $y_{k+1}$ is drawn from the density $p^B$ in $B(x_k,R_k)$.
A term contributes zero when it is not evaluated.
Importantly, the two probabilities need not sum to one: the terms may be selected exclusively, with $\mathbb{P}^{B}=1-\mathbb{P}^{\partial B}$, or the initial condition term may be evaluated at every step, with $\mathbb{P}^{B}=1$.
We derive two random walk strategies corresponding to these choices.
For both strategies, the recursion terminates upon entering the Dirichlet $\varepsilon$-shell and returns $g(\widebar{x}_k,t_k)$, where $\widebar{x}_k$ is the closest point on the Dirichlet boundary.

\subsubsection{Initial Condition at Time Exhaustion}\label{sec:method:time-exhaustion-estimator}
Our first strategy selects either the recursive boundary term or the initial condition term at each step.
The spatial integral of the heat kernel gives the probability that Brownian motion starting at the center $x$ remains inside the ball for at least time $t$.
Equivalently, remaining inside the ball until time $t$ means that the exit time $\eta$ exceeds $t$.
Using the exit time density $H(\eta;R)$, we therefore write the survival probability as
\begin{align}\label{eq:survivalProbability}
    Q(t;R) := \int_{B(x,R)}G^B(x,y,t)\diff y = \int_t^\infty H(\eta;R)\diff\eta.
\end{align}
Consequently, the complementary probability
\begin{align}
    1-Q(t;R)=\int_0^t H(\eta;R)\diff\eta,
\end{align}
is the probability of exiting the ball before $t$.
Accordingly, we choose
\begin{align}
    \mathbb{P}^{\partial B} &= 1-Q(t_k;R_k),
    &
    \mathbb{P}^{B} &= Q(t_k;R_k).
\end{align}
A natural choice for the corresponding conditional densities is
\begin{align}
    p^{\partial B}(x_{k+1},\eta_k) &= \frac{H(\eta_k;R_k)}{|\partial B(x_k,R_k)|[1-Q(t_k;R_k)]},\label{eq:conditionalExitPDF}\\
    p^B(y_{k+1}) &= \frac{G^B(x_k,y_{k+1},t_k)}{Q(t_k;R_k)}.\label{eq:conditionalHeatKernelPDF}
\end{align}
The factor $1/|\partial B(x_k,R_k)|$ in $p^{\partial B}$ follows because the exit location is uniform and independent of the exit time (\cref{eq:PoissonKernelSeparation}).
Substituting these choices into \cref{eq:generalDirichletEstimator} then cancels the kernels and probability weights entirely, yielding
\begin{align}\label{eq:terminalInitialConditionEstimator}
    \widehat{u}(x_k,t_k) =
    \begin{cases}
        \widehat{u}(x_{k+1},t_k-\eta_k),& \eta_k\leq t_k,\\
        u_0(y_{k+1}),& \eta_k>t_k.
    \end{cases}
\end{align}
Although $Q$ and $1-Q$ define the branch probabilities, they no longer need to be evaluated: sampling $\eta_k$ from the full density $H(\cdot;R_k)$ selects between the two events automatically.
If $\eta_k\leq t_k$, the walk exits at a uniformly sampled point $x_{k+1}$ and continues with remaining time $t_k-\eta_k$.
Otherwise, it runs out of time inside the ball, samples $y_{k+1}$ from the time-conditioned heat kernel in \cref{eq:conditionalHeatKernelPDF}, and evaluates $u_0(y_{k+1})$.
The complete procedure is given in \cref{alg:wos}.

\subsubsection{Initial Condition Connections}\label{sec:method:initial-condition-connections}
\begin{figure}[t]
    \centering
    \includegraphics[width=\linewidth]{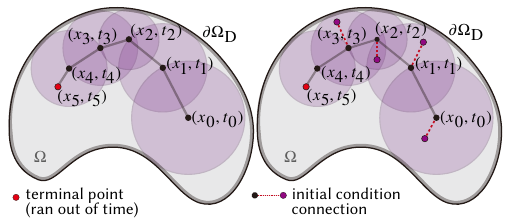}
    \caption{While the straightforward estimation strategy is to sample the initial condition only once inside the domain when a walk runs out of time (\emph{left:} at $x_5$ when $t_5=0$), a more advanced option is to perform an initial condition connection with each walk step (\emph{right}), analogous to how standard walk on spheres samples the source term at a random location inside every ball.}
    \label{fig:concept-initial-condition-connection}
\end{figure}

Alternatively, we retain $\mathbb{P}^{\partial B}=1-Q(t_k;R_k)$ and the conditional densities in \cref{eq:conditionalExitPDF,eq:conditionalHeatKernelPDF}, but set $\mathbb{P}^{B}=1$ so that the initial condition term is evaluated at every step (\cref{fig:concept-initial-condition-connection}).
The boundary weight cancels as before, while sampling $y_{k+1}$ from \cref{eq:conditionalHeatKernelPDF} yields the contribution $Q(t_k;R_k)u_0(y_{k+1})$.
We refer to these per-step evaluations as \emph{initial condition connections}, analogous to per-step source sampling in standard WoS (\cref{sec:bg:wos}) and next-event estimation in Monte Carlo rendering \citep{Veach:1995:Optimally}.
To implement the boundary term selection, we draw $\eta_k$ from the full exit time density $H(\cdot;R_k)$ and continue the walk only if $\eta_k\leq t_k$, an event with probability $1-Q(t_k;R_k)$:
\begin{align}\label{eq:initialConditionConnectionEstimator}
    \widehat{u}(x_k,t_k) = Q(t_k;R_k)u_0(y_{k+1}) +
    \begin{cases}
        \widehat{u}(x_{k+1},t_k-\eta_k),& \eta_k\leq t_k,\\
        0,& \eta_k>t_k.
    \end{cases}
\end{align}
Thus, a single walk accumulates initial condition contributions from every visited ball.

Unlike the first strategy, initial condition connections require evaluating $Q$ explicitly.
In three dimensions, using the dimensionless quantity for time with unit diffusivity, $\tau=t/R^2$, gives
\begin{align}\label{eq:integratedHeatKernel3D}
    Q(t;R) = 2\sum_{n=1}^{\infty}(-1)^{n+1}\exp\left(-n^2\pi^2\tau\right).
\end{align}
In the short-time regime, $\tau\leq0.01$, we approximate $Q$ by one; otherwise, we truncate \cref{eq:integratedHeatKernel3D} after 16 terms.
Both approximations incur relative errors on the order of $10^{-10}$.

In practice, initial condition connections are typically most effective for localized initial conditions at intermediate times (\cref{sec:results:implementation}).

\begin{algorithm*}
\caption{\textsc{WalkOnSpheres}$(x,\heatadd{t},\varepsilon)$\\
\textbf{Note}: Additive changes to steady-state WoS (\cref{sec:bg:wos}) are highlighted in \heatadd{purple}.}
\label{alg:wos}
\begin{algorithmic}[1]
\Require Starting position $x \in \Omega$ of random walk, \heatadd{time budget $t>0$}, and $\varepsilon$-shell.
\Ensure Single-sample estimate $\hat{u}(x\heatadd{,t})$ of heat equation with spatially varying initial condition, source, and Dirichlet data.

\State $d, \widebar{x} \gets \textsc{ClosestPt}(\partial \Omega_{\mathrm{D}}, x)$
\Comment{Compute distance to Dirichlet boundary $\partial \Omega_{\mathrm{D}}$}

\InlineIf{$d < \varepsilon$}{\Return $g(\widebar{x}\heatadd{,t})$}
\Comment{Return value at closest Dirichlet point $\widebar{x}$ \heatadd{and time $t$} if $x$ is inside $\varepsilon$-shell}

\State $v \gets \textsc{SampleUnitSphere}()$
\Comment{Sample direction $v$ uniformly on unit sphere}

\State $p \gets x + d v$
\Comment{Set next walk position on sphere with radius $d$}

\State \heatadd{$\eta \gets \textsc{SampleExitTime}(d)$}
\Comment{\heatadd{Sample exit time from ball of radius $d$ (Sec.~\ref{sec:method:exit-time-sampling})}}

\State $\hat{I}_f \gets \textsc{SourceEstimate}(x,d,\heatadd{t})$
\Comment{Estimate source contribution \heatadd{with time dependence $t$ (\cref{sec:method:inhomogeneous:source})}}

\global\def\algorithmicif{\heatadd{\textbf{if}}}
\global\def\algorithmicthen{\heatadd{\textbf{then}}}
\global\def\algorithmicend{\heatadd{\textbf{end}}}

\InlineIf{\heatadd{$\eta > t$}}{\heatadd{\Return $\textsc{InitialConditionEstimate}(x,d,t) + \hat{I}_f$}}
\Comment{\heatadd{Return total walk contribution if time budget is exhausted (\cref{sec:method:heat-kernel-sampling})}}

\global\def\algorithmicif{\textbf{if}}
\global\def\algorithmicthen{\textbf{then}}
\global\def\algorithmicend{\textbf{end}}

\State \Return $\textsc{WalkOnSpheres}(p,\heatadd{t-\eta},\varepsilon)
    +\hat{I}_f$
    \Comment{Continue from next walk position $p$\heatadd{with remaining time budget $t-\eta$}}

\end{algorithmic}
\end{algorithm*}

\subsection{Exit Time and Heat Kernel Sampling}
\label{sec:method:sampling}

The preceding estimators require sampling exit times from $H$ (\cref{eq:PoissonKernelSeparation}) and initial condition locations from the time-conditioned heat kernel density $G^B/Q$ (\cref{eq:HeatKernel3D,eq:integratedHeatKernel3D}).
We next develop efficient three-dimensional samplers for both.

\subsubsection{Exit Time Sampling}\label{sec:method:exit-time-sampling}

As observed by \citet{HajiSheikh:1966:Floating}, the rescaled exit time density is independent of the ball radius when expressed using dimensionless time $\tau$:
\begin{align}\label{eq:exitTimeDensityScaling3D}
    H(\eta;R)=\frac{1}{R^2}H(\tau;1).
\end{align}
This scaling lets us use the same unit-ball sampler at every walk step: draw $\tau\sim H(\cdot;1)$ and return $\eta=R^2\tau$.

The spectral representation of $H$ does not directly suggest how to sample it.
However, a more useful structure emerges from its Laplace transform, which factorizes for the three dimensional unit ball as \citep{Ciesielski:1962:first,Borodin:2002:handbook}
\begin{equation}\label{eq:exitTimeLaplaceProduct3D}
    \mathcal{L}\{H(\cdot;1)\}(s) = \frac{\sqrt{s}}{\sinh\sqrt{s}} = \prod_{n=1}^{\infty}\frac{\lambda_n}{s+\lambda_n},
\end{equation}
where $\lambda_n=n^2\pi^2$.
The inverse Laplace transform then converts this product into a convolution, with each factor $\lambda_n/(s+\lambda_n)$ corresponding to the exponential PDF $h_n(\tau)=\lambda_n \exp{(-\lambda_n\tau)}$ for $\tau\geq0$.
Thus, $H(\cdot;1)$ is an infinite convolution of exponential PDFs \citep{Kent:1980:Eigenvalue}.

A convolution of PDFs is the PDF of the sum of independent random variables drawn from them \citep[Sec.~I.4.4]{Devroye:1986:NonUniform}.
Hence, we can sample from $H(\cdot;1)$, in principle, by independently drawing $\tau_n \sim h_n$ and returning
\begin{align}
    \tau=\sum_{n=1}^{\infty}\tau_n.
\end{align}
Of course, drawing infinitely many variables is impractical.
The $n$th exponential variable has mean $1/\lambda_n=1/(n^2\pi^2)$ and variance $1/\lambda_n^2=1/(n^4\pi^4)$, so later variables contribute less.
Since the Gamma family includes sums of identically distributed exponential variables, it provides a natural approximation for the tail.
We therefore sample the first $K$ exponential variables exactly and approximate the remaining sum with a single Gamma variable:
\begin{equation}\label{eq:exitTimeApproximation}
    \tau \approx \sum_{n=1}^{K}\tau_n + G,
\end{equation}
where $G\sim\operatorname{Gamma}(\alpha,\theta)$ has shape $\alpha$ and scale $\theta$.

We choose $\alpha$ and $\theta$ so that $G$ matches the exact aggregate mean and variance of the remaining exponential variables.
Since the variables $\tau_n$ are independent, their means and variances add.
Subtracting the first $K$ terms from those of $\tau=\sum_{n=1}^{\infty}\tau_n$ gives
\begin{equation}\label{eq:exitTimeRemainderMoments3D}
    \mu_G=\frac{1}{6}-\sum_{n=1}^{K}\lambda_n^{-1},\qquad \sigma_G^2=\frac{1}{90}-\sum_{n=1}^{K}\lambda_n^{-2}.
\end{equation}
Here, the constants follow from the standard identities $\sum_{n=1}^{\infty}n^{-2}=\pi^2/6$ and $\sum_{n=1}^{\infty}n^{-4}=\pi^4/90$.
A Gamma variable has mean $\alpha\theta$ and variance $\alpha\theta^2$, so to match the target moments, we require $\alpha\theta=\mu_G$ and $\alpha\theta^2=\sigma_G^2$.
Solving these two equations then gives
\begin{align}\label{eq:exitTimeGammaParameters3D}
    \alpha &= \frac{\mu_G^2}{\sigma_G^2},
    &
    \theta &= \frac{\sigma_G^2}{\mu_G}.
\end{align}

\begin{figure*}[t]
    \centering
    \includegraphics[width=\linewidth]{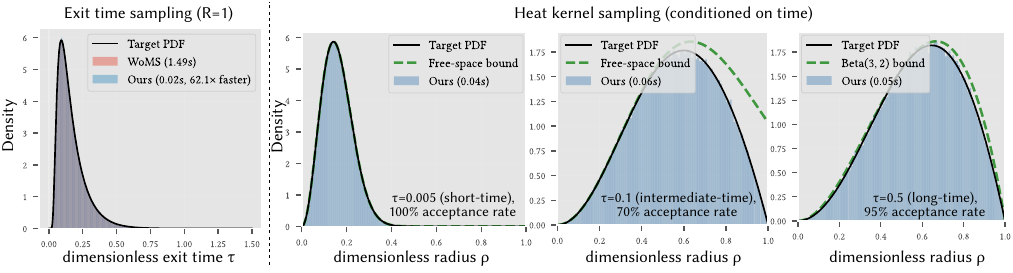}
    \caption{Our tabulation-free exit time sampler for the ball (\emph{left}) accurately reproduces the target exit time distribution (here by generating 200k samples), while being an order of magnitude faster than the walk on moving spheres (WoMS) method of \citet{Deaconu:2017:Walk}. Likewise, our rejection-based radial sampler for the time-conditioned heat kernel (\emph{right}) has low sample rejection and high efficiency across short-, intermediate-, and long-time regimes.}
    \label{fig:exit-time-heat-kernel-sampling}
\end{figure*}

Putting these pieces together, our sampler independently draws the first $K$ variables $\tau_n$ from their exponential densities, draws $G$ from the moment-matched Gamma distribution, and returns
\begin{equation}\label{eq:exitTimeSampler3D}
    \eta = R^2\left(\sum_{n=1}^{K}\tau_n+G\right).
\end{equation}
We use $K=5$ in all experiments.
The sampler exactly preserves the exit time's mean and variance; its Gamma approximation introduces small higher-moment errors, with a maximum relative error of $5.8\times10^{-6}$ over moments $3$ through $12$ in three dimensions.
For comparison, following \citet{HajiSheikh:1966:Floating}, numerical inversion of a 1000-entry CDF table produces an error of $5.6\times10^{-3}$, roughly $1000$ times larger.
Our sampler requires no tabulation and is also substantially faster than WoMS (\cref{fig:exit-time-heat-kernel-sampling}, \emph{left}).
We provide the analogous two dimensional construction in the supplemental (S.2).

\subsubsection{Heat Kernel Sampling}\label{sec:method:heat-kernel-sampling}

Evaluating the initial condition integral, either when a walk runs out of time or through an initial condition connection (\cref{sec:method:initial}), requires sampling an interior point $y$ from the time-conditioned heat kernel density in \cref{eq:conditionalHeatKernelPDF}.
Because the ball $B(x, R)$ is centered at $x$, this density is radially symmetric.
Writing $y=x+R\rho\omega$, the direction $\omega\in\mathbb{S}^2$ is uniformly distributed and the dimensionless radius $\rho=r/R\in[0,1]$ has density
\begin{equation}\label{eq:conditionalHeatKernelRadialDensity3D}
    p^B(\rho;\tau) = \frac{4\pi R^3\rho^2G^B(x,y,t)}{Q(t;R)}.
\end{equation}
Here, the factor $4\pi R^3\rho^2$ accounts for the spherical coordinate Jacobian, the change from $r$ to $\rho$, and integration over direction.
We therefore draw $\omega$ uniformly on the unit sphere and $\rho$ from $p^B$, then return $y=x+R\rho\omega$.
The remaining challenge is to sample $\rho$ efficiently across multiple time scales.

\paragraph{Stable Kernel Evaluation}
The spectral representation of $G^B$ in \cref{eq:HeatKernel3D} converges rapidly at long times but requires many terms and suffers from floating-point cancellation at short times.
For $\tau\leq0.01$, we instead use a method-of-images approximation \citep{Kac:1966:Can}
\begin{equation}\label{eq:shortTimeHeatKernel3D}
    G^B(x,y,t) \approx G^{\mathrm{free}}(r,t) - \frac{2R-r}{r} G^{\mathrm{free}}(2R-r,t),
\end{equation}
where the three dimensional free-space heat kernel is
\begin{equation}
    G^{\mathrm{free}}(r,t) = \frac{1}{(4\pi t)^{3/2}}\exp\left(-\frac{r^2}{4t}\right).
\end{equation}
For longer times, $\tau>0.01$, we truncate the spectral representation after 16 terms.
Together, these representations provide stable kernel evaluations across time scales.

We sample the resulting radial densities using rejection sampling.
Importantly, rejection sampling requires the target density only up to a normalization constant, so the survival probability $Q(t;R)$ in \cref{eq:conditionalHeatKernelRadialDensity3D} need not be evaluated when drawing $\rho$.

\paragraph{Short Times}
For $\tau\leq0.01$, substituting the method-of-images approximation in \cref{eq:shortTimeHeatKernel3D} into \cref{eq:conditionalHeatKernelRadialDensity3D} gives the radial density up to normalization:
\begin{align}\label{eq:shortTimeRadialDensity3D}
    p^B(\rho;\tau)
    &\propto
    \underbrace{\frac{\rho^2}{2\sqrt{\pi}\tau^{3/2}}\exp\left(-\frac{\rho^2}{4\tau}\right)
    }_{q(\rho;\tau)}
    \underbrace{
        \left[1-\frac{2-\rho}{\rho}\exp\left(-\frac{1-\rho}{\tau}\right)\right]
    }_{M(\rho;\tau)}.
\end{align}
The first factor, $q$, is the radial density of a three-dimensional Gaussian vector $\mathbf{X}\sim\mathcal{N}(\mathbf{0},2\tau I_3)$.
The second factor is at most one but becomes negative in an exponentially small neighborhood of $\rho=0$.
The affected probability mass is negligible, so clamping is numerically insignificant.
Accordingly, we sample $\mathbf{X}$, set $\rho=\lVert\mathbf{X}\rVert$, reject it if $\rho\geq1$, and otherwise accept it with probability $\max\{0,M(\rho;\tau)\}$.
Repeating until acceptance samples from the normalized radial density without evaluating its normalization constant.

\paragraph{Intermediate Times}
For $0.01<\tau\leq0.12$, substituting the 16-term spectral expansion of $G^B$ into \cref{eq:conditionalHeatKernelRadialDensity3D} gives
\begin{equation}\label{eq:spectralRadialDensity3D}
    p^B(\rho;\tau) \propto 2\pi\rho\sum_{n=1}^{16}n\sin(n\pi\rho)\exp(-n^2\pi^2\tau).
\end{equation}
Because $G^B$ is bounded by $G^{\mathrm{free}}$, we reuse the Gaussian radial density $q$ from the short-time regime to generate candidate radii.
After rejecting candidates with $\rho\geq1$, we accept them with probability
\begin{equation}\label{eq:intermediateHeatKernelAcceptance3D}
    \mathbb{P}(\mathrm{accept}\mid\rho) = \frac{2\pi\rho}{q(\rho;\tau)}\sum_{n=1}^{16}n\sin(n\pi\rho)\exp(-n^2\pi^2\tau).
\end{equation}

\paragraph{Long Times}
For $\tau>0.12$, the first spectral mode of $G^B$ dominates, and the normalized radial density $p^B(\rho;\tau)$ approaches $\pi\rho\sin(\pi\rho)$.
A $\operatorname{Beta}(3,2)$ density closely matches this limiting shape.
In the supplemental document (S.4.2), we show that the unnormalized radial density in \cref{eq:spectralRadialDensity3D} is bounded by the tight envelope
$C(\tau)q_\beta(\rho)$, where $q_\beta$ denotes the $\operatorname{Beta}(3,2)$ density
\begin{equation}
    q_\beta(\rho)=12\rho^2(1-\rho),
\end{equation}
and
\begin{equation}\label{eq:longTimeRadialBound3D}
    C(\tau) = \frac{2\pi}{3}\sum_{n=1}^{16}n^2\exp(-n^2\pi^2\tau).
\end{equation}
The resulting acceptance probability is
\begin{equation}\label{eq:longHeatKernelAcceptance3D}
    \mathbb{P}(\mathrm{accept}\mid\rho)=\frac{2\pi\rho}{C(\tau)q_\beta(\rho)}\sum_{n=1}^{16}n\sin(n\pi\rho)\exp(-n^2\pi^2\tau).
\end{equation}
We sample from $q_\beta$ using order statistics: the third-smallest of four independent uniform variables on $[0,1)$ follows a $\operatorname{Beta}(3,2)$ distribution \citep[Sec.~IX.4.2]{Devroye:1986:NonUniform}.
We then accept the resulting candidate with the probability in \cref{eq:longHeatKernelAcceptance3D}.

These rejection samplers introduce no additional bias beyond the numerical approximations used to evaluate the kernels.
Across their respective time regimes, they maintain low rejection rates and enable efficient radial sampling (\cref{fig:exit-time-heat-kernel-sampling}, \emph{right}).
We provide the corresponding two dimensional kernel evaluations and sampling procedures in the supplemental document (S.4.1).

\subsection{Walk on Stars with General Initial Conditions}
\label{sec:method:wost}

We next extend the preceding formulation to mixed Dirichlet--Neumann problems using WoSt.
Here, we assume $f=h=0$ and defer nonzero source and Neumann data to \cref{sec:method:inhomogeneous}.
At each step, we construct the star-shaped region $\mathrm{St}(x,R)=\Omega\cap B(x,R)$, with $R$ selected exactly as in standard WoSt (\cref{sec:bg:wost}).
Specializing Green's representation formula for the heat equation \citep{Costabel:1990:Boundary} to this region gives
\begin{align}\label{eq:heatStarInitialIntegral}
    \alpha(x)u(x,t) &= \int_0^t\int_{\partial\mathrm{St}(x,R)}P^B(x,z,\eta)u(z,t-\eta)\diff z\diff\eta\notag\\
                    &\quad+\int_{\mathrm{St}(x,R)}G^B(x,y,t)u_0(y)\diff y.
\end{align}
The primary change from the WoS formulation in \cref{sec:method:initial} is that the integrals are restricted to a star-shaped region rather than an enclosing ball.
The heat kernel $G^B(x,y,t)$ retains its interpretation from the ball formulation, providing the density of Brownian motion reaching $y$ after time $t$ without crossing $\partial B(x,R)$.
The time-dependent Poisson kernel $P^B(x,z,\eta):=-\partial G^B(x,z,\eta)/\partial n_z$ describes time-resolved flux through the star boundary $\partial\mathrm{St}$.
On the spherical portion $\partial\mathrm{St}_{\mathrm B}$, $P^B$ is the joint density of the ball exit time and location (\cref{eq:PoissonKernelSeparation}).
On $\partial\mathrm{St}_{\mathrm N}$, however, it measures flux across a surface inside the enclosing ball, not the joint distribution of when and where Brownian motion first leaves $\mathrm{St}$.
This distinction arises because $G^B$ is the heat kernel of $B$, not $\mathrm{St}$---the associated Brownian motion is absorbed only at $\partial B$ and crosses $\partial\mathrm{St}_{\mathrm N}$ without stopping.

\subsubsection{Recursive Walk}\label{sec:method:wost:recursive-walk}
Although $P^B$ lacks a first-exit interpretation on $\partial\mathrm{St}_{\mathrm N}$, the recursive walk only requires sampling the kernel in \cref{eq:heatStarInitialIntegral}, which factorizes into spatial and temporal densities.
In three dimensions,
\begin{align}\label{eq:starPoissonKernelFactorization}
    P^B(x,z,\eta) &= \underbrace{\frac{n_z\cdot(z-x)}{4\pi\lVert z-x\rVert^3}}_{P^B(x,z)}H_{\mathrm{St}}(\eta;\rho,R),& \rho&=\frac{\lVert z-x\rVert}{R},
\end{align}
where $P^B(x,z)$ is the steady-state Poisson kernel in \cref{eq:wostIntegral} used by standard WoSt.
The conditional temporal density is
\begin{align}\label{eq:starConditionalTimeDensity3D}
    H_{\mathrm{St}}(\eta;\rho,R) &= \frac{2\pi}{R^2}\sum_{n=1}^{\infty}e_n(\rho)\exp\left(-\frac{n^2\pi^2\eta}{R^2}\right),\\
    \text{where}\quad e_n(\rho) &:= n\left[\sin(n\pi\rho)-n\pi\rho\cos(n\pi\rho)\right].\notag
\end{align}
It integrates to one over $\eta\in[0,\infty)$ and reduces to the ball exit time density $H(\eta;R)$ when $\rho=1$.
Thus, \cref{eq:starPoissonKernelFactorization} generalizes \cref{eq:PoissonKernelSeparation} from points on $\partial B$ to points on $\partial\mathrm{St}$ lying inside the enclosing ball.

\begin{figure}[t]
    \centering
    \includegraphics[width=\linewidth]{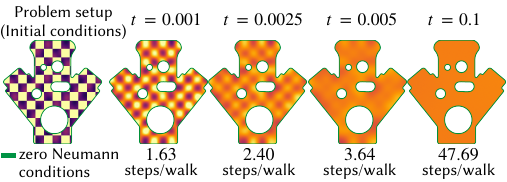}
    \caption{Unlike Laplace problems with pure Neumann boundary conditions, the heat equation with a finite time budget provides an unbiased stopping criterion for our walk on stars estimator---a walk simply terminates inside the domain once it runs out of time. Consequently, shorter walks with smaller time budgets retain the high-frequency details in the initial conditions, while longer walks with larger time budgets capture the more constant, averaged initial condition profile over the domain.}
    \label{fig:pure-neumann-2d}
\end{figure}

The factorization in \cref{eq:starPoissonKernelFactorization} lets us retain the spatial sampling procedure of standard WoSt.
In particular, we exactly importance sample the spatial factor $P^B(x,z)$ by drawing a direction uniformly and intersecting its ray with the star boundary $\partial\mathrm{St}$ \citep{Veach:1995:Optimally}.
When $x_k$ lies on the Neumann boundary, we instead sample from the inward hemisphere, as in standard WoSt \citep[Sec.~4.4.4]{Sawhney:2023:Walk}.
The unique intersection defines the next walk location $x_{k+1}$ and normalized hit distance $\rho_k=\lVert x_{k+1}-x_k\rVert/R_k$.
If the ray reaches $\partial\mathrm{St}_{\mathrm B}$, then $\rho_k=1$, and we reuse the ball exit time sampler from \cref{sec:method:exit-time-sampling} to draw $\eta_k \sim H(\cdot;R_k)$.
If it instead reaches $\partial\mathrm{St}_{\mathrm N}$, then $\rho_k<1$, and we sample $\eta_k \sim H_{\mathrm{St}}(\cdot;\rho_k,R_k)$ using the rejection sampler described in the supplemental document (S.3).
If $\eta_k\leq t_k$, the walk continues from $x_{k+1}$ with remaining time $t_k-\eta_k$; otherwise, the recursion terminates.

The walk returns the prescribed value $g$ upon reaching the Dirichlet $\varepsilon$-shell.
For a pure Neumann problem, no such shell exists, but a sampled elapsed time exceeding the remaining budget still terminates the walk.
The initial condition therefore anchors the time-dependent solution (\cref{fig:pure-neumann-2d}), unlike the corresponding steady-state problem, whose solution is defined only up to an additive constant.

\subsubsection{Initial Condition Contribution}\label{sec:method:wost:initial}
Evaluating the initial condition only at time exhaustion would require sampling a point from the heat kernel conditioned to lie inside $\mathrm{St}(x_k,R_k)$.
Its normalization--the heat kernel integral over this region--depends on local geometry and, unlike the ball integral $Q(t_k;R_k)$ in \cref{eq:survivalProbability}, is not available in closed form.
Instead, we evaluate an initial condition connection at each step (\cref{sec:method:initial-condition-connections}), sampling from the normalized heat kernel of the enclosing ball and extending $u_0$ by zero outside the star region.

\setlength{\columnsep}{0.75em}
\setlength{\intextsep}{0.75em}
\begin{wrapfigure}{r}{0.48\linewidth}
    \vspace{-0.85\baselineskip}
    \centering
    \includegraphics[width=\linewidth]{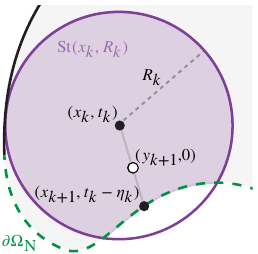}
\end{wrapfigure}
Specifically, we sample $y_{k+1}$ from the density $p^B$ in \cref{eq:conditionalHeatKernelPDF} by reusing the direction $\omega_k$ from the recursive walk and drawing a normalized radial coordinate $\rho_y\in[0,1]$ using \cref{sec:method:heat-kernel-sampling}.
The resulting point is $y_{k+1}=x_k+R_k\rho_y\omega_k$ (inset).
Consequently, $y_{k+1}$ lies inside the star region exactly when $\rho_y\leq\rho_k$, where $\rho_k$ is the normalized distance to the star boundary along $\omega_k$.
The initial condition contribution is therefore
\begin{equation}\label{eq:wostInitialConditionContribution}
    \begin{cases}
        Q(t_k;R_k)u_0(y_{k+1}), & \rho_y\leq\rho_k,\\
        0, & \rho_y>\rho_k.
    \end{cases}
\end{equation}
Assigning zero to samples outside $\mathrm{St}(x_k,R_k)$ preserves an unbiased estimate of the initial condition integral in \cref{eq:heatStarInitialIntegral}.
We summarize the resulting time-dependent WoSt procedure in \cref{alg:wost}.

\begin{figure}[t]
    \centering
    \includegraphics[width=\linewidth]{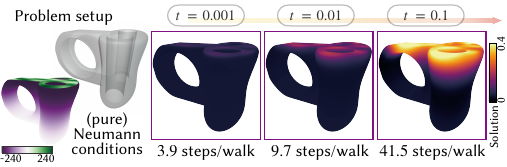}
    \caption{An analogous experiment to \cref{fig:pure-neumann-2d} with nonzero Neumann boundary conditions and zero initial conditions. While smaller time budgets lead to walks terminating quickly, larger time budgets result in long walks that continuously reflect off the boundary. The overhead of exit time sampling is small relative to the cost of performing geometric queries, which remain the primary performance bottleneck for walk on stars.}
    \label{fig:pure-neumann-3d}
\end{figure}

\begin{algorithm*}
\caption{\textsc{WalkOnStars}$(x,n_x,\heatadd{t},\varepsilon)$\\
\textbf{Note}: Additive changes to steady-state WoSt (\cref{sec:bg:wost}) are highlighted in \heatadd{purple}.}
\label{alg:wost}
\begin{algorithmic}[1]
\Require Starting position $x \in \Omega$ of random walk, normal $n_x$ at $x$ (undefined if $x \notin \partial\Omega_{\mathrm{N}}$), \heatadd{time budget $t>0$}, and $\varepsilon$-shell.
\Ensure Single-sample estimate $\hat{u}(x\heatadd{,t})$ of heat equation with spatially varying initial condition, source, and mixed Dirichlet--Neumann data.

\State $d, \widebar{x} \gets \textsc{ClosestPt}(\partial \Omega_{\mathrm{D}}, x)$
\Comment{Compute distance to Dirichlet boundary $\partial \Omega_{\mathrm{D}}$; set $d=\infty$ if $\partial\Omega_{\mathrm D}=\emptyset$}

\InlineIf{$d < \varepsilon$}{\Return $g(\widebar{x}\heatadd{,t})$}
\Comment{Return value at closest Dirichlet point $\widebar{x}$ \heatadd{and time $t$} if $x$ is inside $\varepsilon$-shell}

\State $R \gets \max(\textsc{StarRegionRadius}(\partial \Omega_{\mathrm{N}},x,d), \varepsilon)$
\Comment{Compute radius, $\varepsilon \leq R \leq d$, of star-shaped region}

\State $v \gets \textsc{SampleUnitSphere}()$
\Comment{Sample direction $v$ uniformly on unit sphere}

\InlineIf{$x \in \partial\Omega_{\mathrm{N}}$ \textbf{and} $n_x \cdot v > 0$}{$v \gets -v$}
\Comment{Ensure $v$ is sampled on hemisphere with axis $-n_x$ if $x \in \partial\Omega_{\mathrm{N}}$}

\State $\mathrm{hit}, p, n_p \gets \textsc{Intersect}(\partial\Omega_{\mathrm{N}},x,v,R)$
\Comment{Find first Neumann boundary hit along $x+sv$, $0<s\leq R$}

\InlineIf{\textbf{not} $\mathrm{hit}$}{$p \gets x + Rv$}
\Comment{Set next walk position on $\partial\mathrm{St}_{\mathrm B}$ if $\partial\mathrm{St}_{\mathrm N}$ is not hit; $n_p$ remains undefined}

\State \heatadd{$\eta \gets \textsc{SampleElapsedTime}(x,p,R)$}
\Comment{\heatadd{Sample elapsed time conditioned on boundary point $p$ (Sec.~\ref{sec:method:wost:recursive-walk})}}

\State \heatadd{$\hat{I}_0 \gets \textsc{InitialConditionEstimate}(x,p,v,R,t)$}
\Comment{\heatadd{Estimate initial condition contribution (\cref{sec:method:wost:initial})}}

\State $\hat{I}_f \gets \textsc{SourceEstimate}(x,p,v,R\heatadd{,t})$
\Comment{Estimate source contribution \heatadd{with time dependence $t$ (\cref{sec:method:inhomogeneous:source})}}

\State $\hat{I}_h \gets \textsc{NeumannEstimate}(x,R\heatadd{,t})$
\Comment{Estimate Neumann contribution \heatadd{with time dependence $t$ (\cref{sec:method:inhomogeneous:neumann})}}

\global\def\algorithmicif{\heatadd{\textbf{if}}}
\global\def\algorithmicthen{\heatadd{\textbf{then}}}
\global\def\algorithmicend{\heatadd{\textbf{end}}}

\InlineIf{\heatadd{$\eta > t$}}{\heatadd{\Return $\hat{I}_0 + \hat{I}_h + \hat{I}_f$}}
\Comment{\heatadd{Return total walk contribution if time budget is exhausted}}

\global\def\algorithmicif{\textbf{if}}
\global\def\algorithmicthen{\textbf{then}}
\global\def\algorithmicend{\textbf{end}}

\State \Return $\textsc{WalkOnStars}(p,n_p,\heatadd{t-\eta},\varepsilon)
    + \heatadd{\hat{I}_0} + \hat{I}_h + \hat{I}_f$
    \Comment{Continue from next walk position $p$\heatadd{with remaining time budget $t-\eta$}}

\end{algorithmic}
\end{algorithm*}

\subsection{Time-Dependent Source and Neumann Data}
\label{sec:method:inhomogeneous}

We now complete the representation in \cref{eq:heatStarInitialIntegral} by incorporating general time-dependent source and Neumann data.
Both produce additional space-time integrals evaluated at every random walk step, but require different sampling strategies.
We first consider the source term, reusing the time-conditioned heat kernel sampler developed for initial conditions, before turning to Neumann data.

\subsubsection{Source Term}\label{sec:method:inhomogeneous:source}
A nonzero source $f$ adds the volume integral
\begin{equation}\label{eq:heatSourceIntegralStar}
    \int_0^t\int_{\mathrm{St}(x,R)}G^B(x,y,\eta)f(y,t-\eta)\diff y\diff\eta
\end{equation}
to the right-hand side of \cref{eq:heatStarInitialIntegral}.
For pure Dirichlet problems, $\mathrm{St}(x,R)=B(x,R)$, recovering the corresponding WoS integral.

As with the initial condition contribution, normalizing the space-time heat kernel over the star region would require a geometry-dependent integral.
We instead extend $f$ by zero outside $\mathrm{St}(x,R)$ and sample from the space-time heat kernel of the enclosing ball.
Its normalization is
\begin{equation}\label{eq:sourceHeatKernelNormalization}
    Z(t;R) := \int_0^t\int_{B(x,R)} G^B(x,y,\eta)\diff y\diff\eta = \int_0^t Q(\eta;R)\diff\eta.
\end{equation}
\setlength{\columnsep}{0.75em}
\setlength{\intextsep}{0.75em}
\begin{wrapfigure}{r}{0.48\linewidth}
    \vspace{-0.65\baselineskip}
    \centering
    \includegraphics[width=\linewidth]{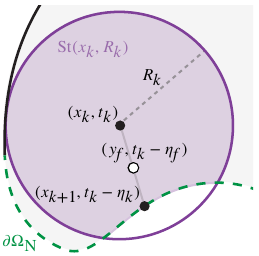}
\end{wrapfigure}
We first sample an elapsed time $\eta_f\in[0,t]$ from the density $Q(\eta_f;R)/Z(t;R)$ using the sampler described in the supplemental document (S.4.3 and S.4.4).
The interior location then has density $p^B$ in \cref{eq:conditionalHeatKernelPDF}, with its time argument set to $\eta_f$.
We sample this location using the procedure in \cref{sec:method:heat-kernel-sampling}.

\begin{figure}[t]
    \centering
    \includegraphics[width=\linewidth]{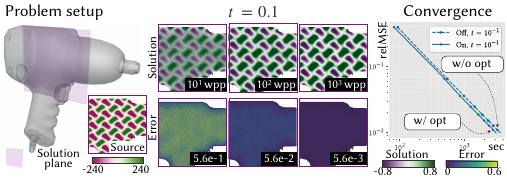}
    \caption{Our method exhibits predictable convergence on a source-dominated manufactured solution test as the number of walks per point increases, shown here at $t=0.1$. The source is derived from the prescribed solution as $f=\partial u/\partial t-\kappa\Delta u$. In this example, the time-independent source optimization (opt) from \cref{sec:efficiency:time-independent} improves efficiency by $15\%$.}
    \label{fig:variance-source}
\end{figure}

For WoSt, we reuse the direction $\omega_k$ sampled by the recursive walk and draw a normalized radial coordinate $\rho_f\in[0,1]$ conditioned on $\eta_f$.
This gives $y_f=x_k+R_k\rho_f\omega_k$, as shown in the inset.
The source contribution at step $k$ is
\begin{equation}\label{eq:sourceContributionStar}
    \begin{cases}
        Z(t_k;R_k)f(y_f,t_k-\eta_f),& \rho_f\leq\rho_k,\\
        0,& \rho_f>\rho_k.
    \end{cases}
\end{equation}
For WoS, $\rho_k=1$, so every sampled source location lies inside the ball and contributes.
The resulting estimator exhibits predictable Monte Carlo convergence on source-dominated problems (\cref{fig:variance-source}).

\subsubsection{Neumann Boundary Condition}\label{sec:method:inhomogeneous:neumann}
A nonzero Neumann condition $h$ further adds the boundary integral
\begin{equation}\label{eq:heatNeumannIntegralStar}
    \int_0^t\int_{\partial\mathrm{St}_{\mathrm N}(x,R)}G^B(x,z,\eta)h(z,t-\eta)\diff z\diff\eta
\end{equation}
\setlength{\columnsep}{0.75em}
\setlength{\intextsep}{0.75em}
\begin{wrapfigure}[11]{r}{0.48\linewidth}
    \vspace{-0.65\baselineskip}
    \centering
    \includegraphics[width=\linewidth]{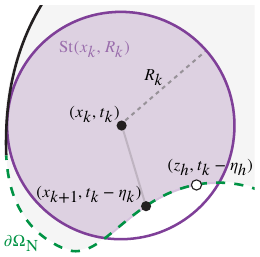}
\end{wrapfigure}
to the right-hand side of \cref{eq:heatStarInitialIntegral}.
Here, $h$ replaces the normal derivative of $u$ on $\partial\mathrm{St}_{\mathrm N}$, while the corresponding integral over $\partial\mathrm{St}_{\mathrm B}$ vanishes because $G^B=0$ on $\partial B(x,R)$.

We estimate \cref{eq:heatNeumannIntegralStar} by first sampling a boundary location and then an elapsed time conditioned on that location (inset).
In particular, at step $k$, we sample $z_h$ on $\partial\mathrm{St}_{\mathrm N}(x_k,R_k)$ from a surface density $p^{\partial\mathrm{St}_{\mathrm N}}$, using the hierarchical importance sampling procedure of standard WoSt \citep[Sec.~4.5]{Sawhney:2023:Walk}.
Once $z_h$ has been selected, we draw $\eta_h\in[0,t_k]$ proportionally to $G^B(x_k,z_h,\eta_h)$.
The required normalization is the time-integrated heat kernel
\begin{equation}\label{eq:residenceDensity}
    W(x,z;t):=\int_0^t G^B(x,z,\eta)\diff\eta.
\end{equation}
The supplemental provides procedures for evaluating $W$ and sampling the resulting density in two and three dimensions (S.5--6).

Using these spatial and temporal densities, $G^B$ cancels in the Monte Carlo estimator of \cref{eq:heatNeumannIntegralStar}.
Dividing by the coefficient $\alpha(x_k)$ on the left-hand side of \cref{eq:heatStarInitialIntegral} then yields the contribution
\begin{equation}\label{eq:neumannContributionStar}
    \frac{W(x_k,z_h;t_k)h(z_h,t_k-\eta_h)}{\alpha(x_k)p^{\partial\mathrm{St}_{\mathrm N}}(z_h)}.
\end{equation}
Accumulating this contribution at every walk step supports spatially and temporally varying Neumann data.
\Cref{fig:pure-neumann-3d} demonstrates the resulting estimator on a pure Neumann problem.

\subsection{Constant Non-Unit Diffusivity}
\label{sec:method:diffusivity}

Thus far, we have assumed unit diffusivity.
Any positive constant diffusivity $\kappa$ can be handled by rescaling time.
Define $\widehat{t}=\kappa t$ and $v(x,\widehat{t})=u(x,\widehat{t}/\kappa)$.
The transformed solution $v$ then satisfies
\begin{equation}
    \frac{\partial v}{\partial\widehat{t}}(x,\widehat{t}) = \Delta v(x,\widehat{t})+\widehat{f}(x,\widehat{t}),
\end{equation}
with transformed data
\begin{align}
    \widehat{f}(x,\widehat{t}) &= \frac{1}{\kappa}f\left(x,\frac{\widehat{t}}{\kappa}\right),&
    \widehat{g}(x,\widehat{t}) &= g\left(x,\frac{\widehat{t}}{\kappa}\right),\notag\\
    \widehat{h}(x,\widehat{t}) &= h\left(x,\frac{\widehat{t}}{\kappa}\right),&
    \widehat{u}_0(x) &= u_0(x).
\end{align}
The source acquires the factor $1/\kappa$ from the rescaled time derivative; the boundary data are reparameterized in time, while the initial condition remains unchanged.
Consequently, we evaluate the original solution $u(x,t)=v(x,\kappa t)$ by running the unit-diffusivity solver with initial time budget $\widehat{t}_0=\kappa t$.
A sampled elapsed time $\widehat{\eta}$ corresponds to $\eta=\widehat{\eta}/\kappa$ in the original time variable, while the local dimensionless time is $\tau=\widehat{\eta}/R^2=\kappa\eta/R^2$.
No changes to the spatial sampling procedures or geometric queries are required.

\section{Improving Efficiency}
\label{sec:efficiency}

Beyond the efficient kernel samplers developed in \cref{sec:method}, we further improve efficiency by skipping zero initial condition contributions, combining sampling strategies for localized initial conditions, omitting temporal samples for time-independent source and Neumann data, and sharing walks across target times.

\subsection{Localized Initial Conditions}
\label{sec:efficiency:localized}

Sampling from the heat kernel density $p^B$ in \cref{eq:conditionalHeatKernelPDF} is effective for broadly supported initial conditions but often misses localized support.
Before sampling, we therefore test whether a bounding volume of the support intersects the current ball $B(x_k,R_k)$.
If they are disjoint, the initial condition integral is zero and we skip its evaluation.
This inexpensive test is especially useful for initial condition connections, which evaluate the integral at every walk step.

\begin{figure}[t]
    \centering
    \includegraphics[width=\linewidth]{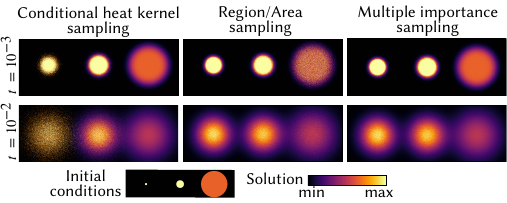}
    \caption{As the time budget and spatial support of the initial condition vary, estimating its contribution exclusively with either heat kernel sampling (\emph{left}) or region sampling (\emph{middle}) can become inefficient. We instead combine proposals from both sampling routines using multiple importance sampling (\emph{right}), enabling robust estimation across time horizons and support sizes.}
    \label{fig:variance-mis}
\end{figure}

When the volumes overlap, we combine heat kernel sampling with \emph{region sampling} over the nonzero region $A$ of $u_0$.
For example, uniform region sampling uses $p^A(y)=1/|A|$ for $y\in A$ and zero otherwise.
We combine these strategies using multiple importance sampling (MIS) with the balance heuristic \citep{Veach:1997:Robust,Pharr:2016:Physically}.
With one sample $y_A$ from $p^A$ and one sample $y_B$ from $p^B$ (\cref{sec:method:heat-kernel-sampling}), the combined contribution is
\begin{align}\label{eq:initialConditionMIS}
    w_A(y_A)\frac{G^B(x_k,y_A,t_k)u_0(y_A)}{p^A(y_A)} + w_B(y_B)\frac{G^B(x_k,y_B,t_k)u_0(y_B)}{p^B(y_B)},
\end{align}
with weights
\begin{align}\label{eq:initialConditionMISWeights}
    w_A(y) &= \frac{p^A(y)}{p^A(y)+p^B(y)},&
    w_B(y) &= \frac{p^B(y)}{p^A(y)+p^B(y)}.
\end{align}
The heat kernel proposal follows the spatial decay of $G^B$, while the region proposal samples where $u_0$ is nonzero.
Their combination is effective across time budgets and support sizes (\cref{fig:variance-mis}).

\subsection{Time-Independent Source and Neumann Data}
\label{sec:efficiency:time-independent}

Time-independent source and Neumann data require no elapsed time sampling.
Using the time-integrated heat kernel $W$ from \cref{eq:residenceDensity} and writing $\mathrm{St}=\mathrm{St}(x,R)$, the corresponding integrals reduce to
\begin{align}
    \int_0^t\int_{\mathrm{St}}G^B(x,y,\eta)f(y)\diff y\diff\eta &= \int_{\mathrm{St}}W(x,y;t)f(y)\diff y,\label{eq:timeIndependentSource}\\
    \int_0^t\int_{\partial\mathrm{St}_{\mathrm N}}G^B(x,z,\eta)h(z)\diff z\diff\eta &= \int_{\partial\mathrm{St}_{\mathrm N}}W(x,z;t)h(z)\diff z.\label{eq:timeIndependentNeumann}
\end{align}
For the source contribution, we sample $y$ directly from the density $W(x,y;t)/Z(t;R)$ over the enclosing ball and extend $f$ by zero outside $\mathrm{St}$, eliminating the $\eta_f$ sample in \cref{eq:sourceContributionStar}.
For the Neumann contribution, we reuse the hierarchical boundary sampler from \cref{sec:method:inhomogeneous:neumann} and evaluate $W$ at the sampled point, eliminating the $\eta_h$ sample in \cref{eq:neumannContributionStar}.
Because elapsed time does not affect $f$ or $h$, omitting these samples preserves each estimator’s expectation and variance while reducing cost.

\subsection{Shared Walks Across Multiple Times}
\label{sec:efficiency:shared}

 Estimating several target times independently can be expensive since each requires a separate spatial walk.
The spatial steps and exit time samples from one walk can instead be reused across different initial time budgets, allowing it to estimate all requested times.
Conceptually, this resembles reusing a light path across color or wavelength channels in RGB and spectral rendering.

For each target time $t^{j}$, we initialize a budget $t_0^{j}=t^{j}$.
At step $k$, all unfinished estimates reuse the exit location $x_{k+1}$ and exit time $\eta_k$.
If $\eta_k\leq t_k^{j}$, the corresponding estimate continues with budget $t_{k+1}^{j}=t_k^{j}-\eta_k$; otherwise, its budget is exhausted and the estimate terminates.
Initial condition, source, and Neumann contributions are evaluated separately for each target because they depend on its remaining budget.
If the shared walk reaches the Dirichlet boundary, all unfinished estimates evaluate the boundary data at their respective remaining times.
The walk ends upon reaching the Dirichlet boundary or once all target budgets are exhausted.

\begin{figure}[t]
    \centering
    \includegraphics[width=\linewidth]{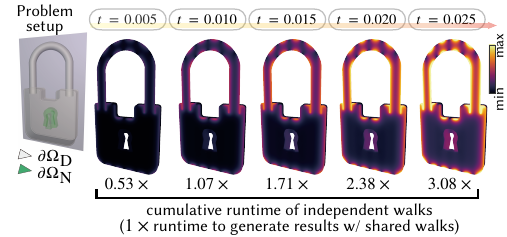}
    \caption{Sharing walks across five different time budgets provides a $3\times$ runtime-efficiency improvement in this example with time-dependent Dirichlet and Neumann boundary conditions.}
    \label{fig:variance-shared-walks}
\end{figure}

Sharing a walk correlates estimates across target times but introduces no additional bias.
More importantly, it amortizes geometric queries across all targets, improving efficiency (\cref{fig:teaser,fig:variance-shared-walks}).

\section{Evaluation}
\label{sec:results}

We first describe our implementation, then compare against alternative grid-free Monte Carlo methods and the finite element method.
We conclude with a representative transient heat conduction problem based on a nuclear fusion reactor component.

\subsection{Implementation}
\label{sec:results:implementation}

We implement our method on the CPU by extending the WoS and WoSt solvers in \textsc{Zombie} \citep{Sawhney:2023:Zombie}.
All CPU experiments use 64 TBB threads on an AMD EPYC 9684X processor.
Following \cref{alg:wos,alg:wost}, we extend the prescribed-data callback routines with a time parameter, add an initial condition callback, and encapsulate time-dependent kernel evaluation and sampling in a reusable utility.
Unless noted otherwise, we set $\varepsilon=0.001L$, where $L$ is the domain's bounding-box diagonal.

We skip initial condition, source, and Neumann contributions when absent and apply the time-independent optimizations from \cref{sec:efficiency:time-independent} whenever possible.
Steady-state problems continue to use the original \textsc{Zombie} estimators.
Shared walk evaluation is disabled by default, except in \cref{fig:teaser,fig:variance-shared-walks}, so other reported results estimate each target time independently.

\begin{figure}[t]
    \centering
    \includegraphics[width=\linewidth]{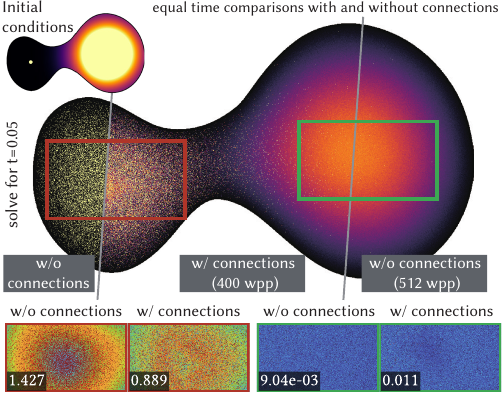}
    \caption{An equal-runtime comparison with and without initial condition connections for a pure Dirichlet heat equation. At the intermediate time shown, connections reduce noise for the localized initial condition (\emph{red inset, right}) despite fewer walks per point. For the broadly supported initial condition (\emph{green inset}), their overhead outweighs the variance reduction.}
    \label{fig:variance-initial-condition-connection}
\end{figure}

Initial condition locations are sampled from the conditional heat kernel unless stated otherwise.
WoSt requires initial condition connections, whereas for pure Dirichlet WoS we enable them only when their variance reduction offsets the additional per-step sampling cost.
This trade-off typically favors connections at intermediate times with localized initial conditions (\cref{fig:variance-initial-condition-connection}).

Geometry queries dominate runtime.
At long times, time-dependent kernel sampling adds modest overhead relative to steady-state WoS and WoSt; at short times, early time exhaustion often makes transient solves faster (\cref{fig:comparison-steady-state-transient-dirichlet,fig:comparison-steady-state-transient-mixed}).

\subsection{Alternative Grid-Free Monte Carlo Solvers}
\label{sec:results:mc-comparisons}

\begin{figure*}[t]
    \centering
    \includegraphics[width=\linewidth]{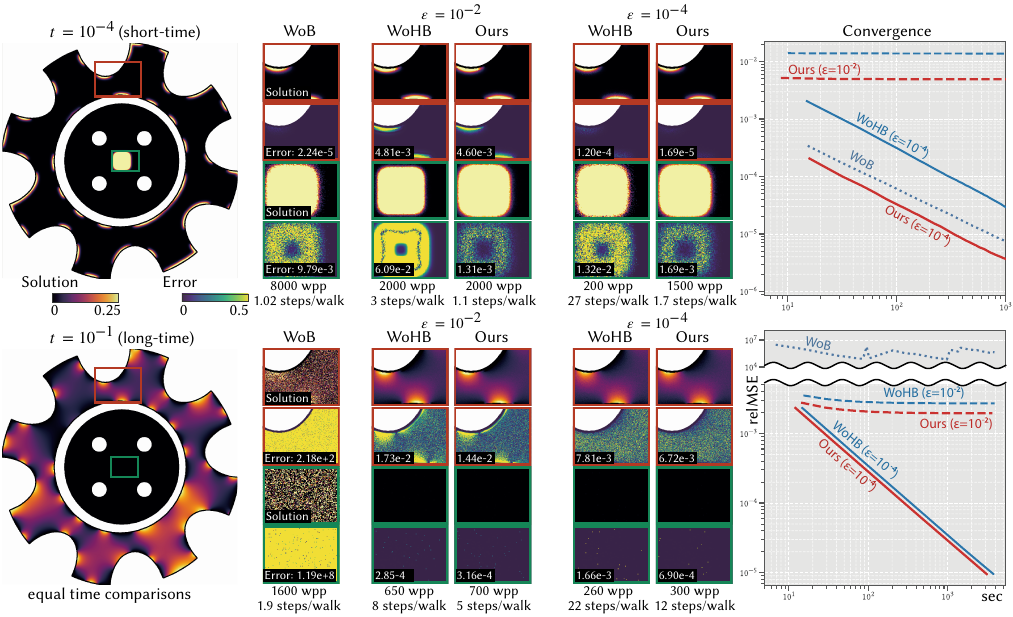}
    \caption{An equal-runtime comparison of walk on boundary (WoB) \citep{Sugimoto:2024:Velocitybased}, walk on heat balls (WoHB) \citep{Deaconu:2018:Initial}, and our method for a heat equation with pure Dirichlet conditions. Although WoB runs roughly $5\times$ more walks per point than our method, its variance becomes extremely high at longer time budgets in non-convex domains (\emph{bottom row}). WoHB and our method instead exhibit predictable Monte Carlo convergence at all time budgets. However, WoHB's spacetime heat balls require smaller spatial steps and epsilon-shell termination near both the boundary and initial time. Consequently, our method has less bias and is $3\times$ faster for the smaller time budget in this example (\emph{top row}).}
    \label{fig:comparison-wohb-wob}
\end{figure*}

We compare against walk on boundary (WoB) and walk on heat balls (WoHB) on the pure Dirichlet problem in \cref{fig:comparison-wohb-wob}.

\subsubsection{Walk on Boundary}
\label{sec:results:wob}

For $f=0$, WoB represents the solution using a double-layer boundary potential and the free-space heat kernel \citep{Sabelfeld:1994:Random,Sugimoto:2024:Velocitybased}:
\begin{align}\label{eq:wobHeatRepresentation}
    u(x,t) &= -\int_0^t\int_{\partial\Omega}\frac{\partial G^{\mathrm{free}}}{\partial n_z}(x,z,t-\eta)\phi(z,\eta)\diff z\diff\eta\notag\\
    &\quad + \int_\Omega G^{\mathrm{free}}(x,y,t)u_0(y)\diff y,
\end{align}
where $\phi$ is an unknown boundary density satisfying a recursive boundary integral equation.
Unlike our method, which walks between local spheres and star-shaped regions, WoB uses ray tracing to construct walks between points on the global domain boundary.

On a non-convex domain, a sampled ray may intersect the boundary multiple times.
WoB selects one intersection and compensates by multiplying the path weight by the number of intersections.
These factors compound along a walk and can produce unbounded variance, as also observed for steady-state WoB \citep[Fig.~12]{Miller:2024:Walkin}.
The problem becomes more severe at longer target times because walks typically contain more steps.
Furthermore, the normal-derivative kernel in \cref{eq:wobHeatRepresentation} changes sign, causing substantial cancellation and occasionally large negative estimates.

For the equal-runtime comparison in \cref{fig:comparison-wohb-wob}, we run the WoB implementation on an NVIDIA RTX 3060 Laptop GPU and our method on the CPU configuration described above.
Although WoB traces roughly $5\times$ more walks per point, its sample variance for the larger time budget is approximately $10^7$, compared with $10^{-3}$ for our method.
Our estimates instead exhibit predictable Monte Carlo convergence across both time budgets.

\subsubsection{Walk on Heat Balls}
\label{sec:results:wohb}

WoHB avoids sampling the exit time from a fixed sphere by constructing a spacetime heat ball at each step \citep{Deaconu:2018:Initial}.
It samples an elapsed time using walk on moving spheres \citep{Deaconu:2013:Hitting,Deaconu:2017:Walk}, then determines a corresponding spatial radius and exit location.
This construction couples the temporal and spatial steps.

To remain within the domain and remaining time budget, each heat ball generally has a smaller spatial radius than the largest inscribed sphere used by WoS, and this radius shrinks further as the time budget approaches zero.
WoHB therefore requires more walk steps, whereas our method uses the largest inscribed sphere and samples its exit time independently.
For the smaller time budget in \cref{fig:comparison-wohb-wob}, our method consequently runs approximately $3\times$ faster.

Moreover, a WoHB elapsed time never exhausts the remaining budget exactly.
The walk instead terminates within a temporal $\varepsilon$-shell and evaluates the initial condition at its current location, introducing additional bias that appears as under-diffusion in \cref{fig:comparison-wohb-wob}.
Our exit time samples may exceed the remaining budget, in which case we draw the terminal location from the time-conditioned heat kernel without a temporal $\varepsilon$-shell.
Our formulation also supports initial condition connections and multiple importance sampling for further variance reduction (\cref{fig:variance-initial-condition-connection,fig:variance-mis}).
Both methods reduce Monte Carlo noise predictably as the number of walks increases, but WoHB retains temporal termination bias.

\subsection{Comparison with the Finite Element Method}
\label{sec:results:fem}

Our baseline is continuous Galerkin FEM with piecewise-linear elements, implemented in MFEM \citep{Anderson:2021:MFEM}.
We evaluate both explicit fourth-order Runge--Kutta (RK4) and implicit Backward Euler time integration.
The following experiments isolate errors from spatial and temporal discretization.

FEM requires a tetrahedral volume mesh.
Fast meshers \citep{Si:2015:TetGen,Diazzi:2026:Surface} often assume clean, watertight surfaces and can be brittle on CAD-derived geometry.
Robust alternatives \citep{Hu:2018:TetWild,Hu:2020:fTetWild} may incur substantial time and memory costs or approximate the boundary, potentially merging nearby features or removing seams (\cref{fig:meshingChallenges}).
A successful mesh must also resolve localized interior data.
To isolate this spatial error in \cref{fig:comparison-fem-spatial-aliasing}, we use a small Backward Euler step.
Coarse meshes under-resolve the localized initial condition, while aggressive adaptive refinement increases runtime and memory yet leaves visible aliasing artifacts.
Our method instead operates on the boundary mesh, avoiding the need to resolve localized interior data on a volume mesh.

Time integration introduces a separate error source.
Explicit schemes such as RK4 are conditionally stable, with the maximum step size constrained by the finest spatial resolution \citep{Courant:1928:Partial}.
Mesh refinement can therefore force prohibitively small time steps, as shown in \cref{fig:comparison-fem-temporal-discretization-initial}.
Backward Euler permits larger stable steps, but these introduce numerical diffusion; reducing the step size improves accuracy at greater cost.
The moving-source experiment in \cref{fig:comparison-fem-temporal-aliasing-source} further shows that stability alone is insufficient---even an
unconditionally stable method needs sufficiently small steps to resolve rapidly changing data.
Choosing an integrator and step size requires balancing accuracy, stability, and computational cost.

Our solver evaluates solutions at arbitrary times without time stepping or temporal discretization bias.
With the $\varepsilon$-shell fixed, more walks predictably reduce noise.
Shared walks estimate solutions at target times without computing intermediate states (\cref{sec:efficiency:shared}).

\subsection{Thermal Analysis}
\label{sec:results:tokamak}

Accurate transient temperature prediction is critical for preventing overheating in thermal engineering systems.
ITER, an experimental tokamak fusion reactor \citep{Rebut:1995:ITER}, provides a demanding example: its divertor vertical targets are expected to sustain heat fluxes of $10$--$20\,\mathrm{MW}/\mathrm{m}^2$, up to ten times that on a spacecraft reentering Earth's atmosphere \citep{ITER:2018:Hottest}.
Maintaining safe operating temperatures is particularly important because, near $1600\,\mathrm{K}$, tungsten-based plasma-facing components can undergo significant recrystallization, increasing their susceptibility to cracking under high heat flux \citep{Wang:2024:Consideration}.

As a representative application of our solver (\cref{fig:teaser}), we simulate heat conduction in the inner vertical target from ITER's publicly released tokamak model \citep{Coblentz:2020:Make}.
Following established thermal analyses of divertor monoblocks, we model the temperature evolution using the heat equation \citep{ElMorshedy:2021:Thermal,VanDenKerkhof:2021:Impact}.
Particles escaping the confined plasma deposit heat in a narrow footprint around a strike line on the target surface.
We prescribe this load as a Neumann condition using the Eich profile \citep{Eich:2011:Inter}.
To distribute the heat load, we sweep the strike line vertically \citep{Silburn:2017:Mitigation}, beginning at $t=1.5\,\mathrm{s}$, and superimpose its moving footprint on a background flux of $3\,\mathrm{MW}/\mathrm{m}^2$.

\begin{figure}[t]
    \centering
    \includegraphics[width=\linewidth]{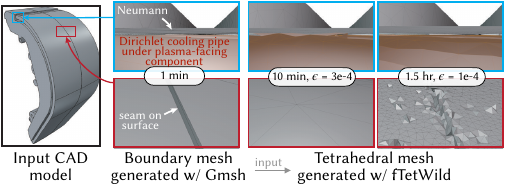}
    \caption{Gmsh triangulates the CAD boundary representation of the ITER vertical target (\cref{fig:teaser}) in under one minute \citep{Remacle:2022:Gmsh}. This boundary mesh suffices for our method, while FEM also requires a tetrahedral volume mesh. Using it as input, fTetWild \citep{Hu:2020:fTetWild} produces a volume mesh whose boundary only approximates the geometry: Dirichlet cooling pipes merge with the Neumann plasma-facing component (\emph{top middle}), and component seams completely disappear (\emph{bottom middle}). A smaller envelope better preserves these features but greatly increases meshing time while introducing severe surface artifacts (\emph{bottom right}).}
    \label{fig:meshingChallenges}
\end{figure}

Cooling pipes are typically modeled with Robin boundary conditions accounting for heat transfer into the coolant.
Because our time-dependent solver does not currently support Robin conditions, we approximate the pipe walls with fixed-temperature Dirichlet conditions.
We also prescribe a nonzero initial temperature throughout the target.
Thus, this experiment is a representative transient thermal analysis rather than a complete thermal-hydraulic model.

Our solver could operate directly on the CAD boundary representation if the required geometric queries were available \citep{Sawhney:2023:Walk}.
Since \textsc{Zombie} currently implements these queries only on triangle meshes \citep{Sawhney:2023:Zombie}, we triangulate the CAD boundary using Gmsh \citep{Remacle:2022:Gmsh}.
The resulting boundary mesh is sufficient for our solver, whereas FEM also requires a tetrahedral volume mesh (\cref{fig:meshingChallenges}).
For the FEM comparison, we generate a volume mesh using fTetWild, refine it adaptively using the Zienkiewicz--Zhu error estimator \citep{Hu:2020:fTetWild,Zienkiewicz:1987:Simple}, and repeat the simulation with smaller time steps to reduce spatial and temporal errors.

Our solver requires neither a volume mesh nor time stepping.
To evaluate temperature near the initial strike-line location, it launches walks only from the selected probe.
Using shared walks (\cref{sec:efficiency:shared}), each walk estimates all $5000$ target times spanning $[0,250]\,\mathrm{s}$.
Using 64 TBB threads, one million shared walks reconstruct this temperature history in about three minutes.
In contrast, FEM advances a global solution sequentially.
With the coarse time step shown in \cref{fig:teaser}, the FEM solve takes about one hour and overestimates the temperature enough to spuriously exceed the safety limit.
Reducing the time step improves the prediction at substantially greater cost.

\begin{figure*}[t]
    \centering
    \includegraphics[width=\linewidth]{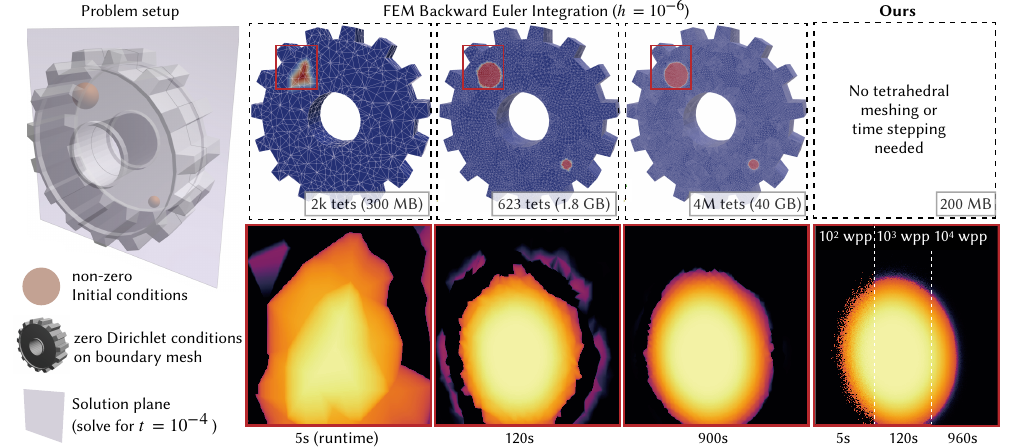}
    \caption{Monte Carlo solution of the heat equation compared with FEM on a simple gear geometry: for this short-time problem ($t=10^{-4}$), FEM on a coarse mesh is fast but suffers from severe spatial aliasing because the localized initial condition is under-resolved (\emph{2nd column}). We use a small Backward Euler step ($h=10^{-6}$) so that spatial discretization dominates the error. Adaptive refinement up to $4$ million tetrahedra greatly increases memory and runtime, yet visible mesh artifacts remain (\emph{3rd \& 4th columns}). The reported FEM runtimes exclude tetrahedral mesh generation, and further refinement fails due to integer overflow. In contrast, our Monte Carlo solver requires neither tetrahedral volume meshing nor time stepping (\emph{5th column}). As a result, it retains a small memory footprint, avoids mesh-induced aliasing, and converges to a consistent heat distribution as the number of walks per point increases.}
    \label{fig:comparison-fem-spatial-aliasing}
\end{figure*}

\begin{figure*}[t]
    \centering
    \includegraphics[width=\linewidth]{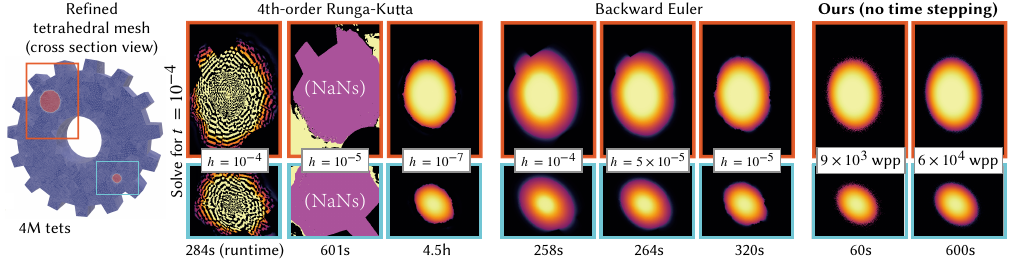}
    \caption{Time stepping introduces an additional source of error for FEM solvers. Explicit high-order integrators such as RK4 are only conditionally stable (\emph{1st block}), with the maximum stable time step $h$ constrained by mesh resolution. On the highly refined mesh used here to reduce spatial aliasing, RK4 requires $h=10^{-7}$ for stability, making the solve prohibitively slow. Backward Euler is unconditionally stable (\emph{2nd block}), but large steps introduce numerical diffusion, while choosing a sufficiently small step requires tuning. In contrast, our Monte Carlo solver avoids time stepping entirely (\emph{3rd  block}), producing estimates with sampling variance rather than time-discretization bias; this variance decreases predictably as more walks are used.}
    \label{fig:comparison-fem-temporal-discretization-initial}
\end{figure*}

\begin{figure*}[t]
    \centering
    \includegraphics[width=\linewidth]{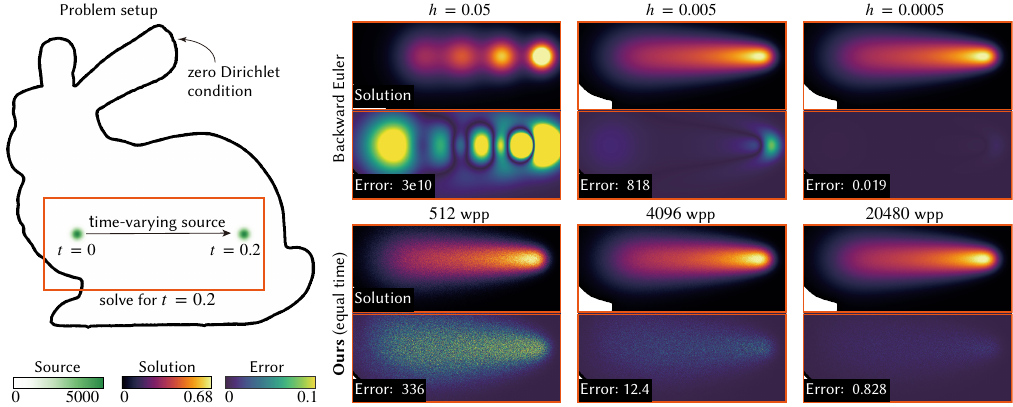}
    \caption{An equal-runtime comparison with FEM for a time-varying source. We solve a 2D heat equation with zero initial and Dirichlet data, diffusivity $\kappa=0.01$, and a localized Gaussian source moving right (\emph{left}). FEM uses Backward Euler on a highly refined 234k-node mesh to minimize spatial error (\emph{top row}). Although unconditionally stable, large time steps under-resolve the source motion and produce substantial temporal error; reducing the step size $h$ recovers the correct heat profile at greater cost. Our Monte Carlo solver instead requires no time stepping and evaluates the solution directly at the target time (\emph{bottom row}), avoiding temporal aliasing. Even with few walks, it captures the overall heat distribution; increasing walks per point predictably reduces sampling noise.}
    \label{fig:comparison-fem-temporal-aliasing-source}
\end{figure*}

\section{Conclusion and Future Work}
\label{sec:conclusion}

We developed time-dependent extensions of WoS and WoSt for IBVPs governed by the heat equation, with general initial conditions and time-dependent source, Dirichlet, and Neumann data.
Our low-bias, tabulation-free exit time sampler and numerically accurate rejection samplers make the required time-dependent kernels practical.
The solver evaluates arbitrary spacetime queries without volumetric meshing or time stepping while retaining the parallel, progressive, and output-sensitive properties of WoS and WoSt.
Shared walks enable efficient evaluation at multiple target times.

The primary limitation remains Monte Carlo variance, particularly at low walk counts.
Neumann-dominated problems with large time budgets can also be expensive because walks repeatedly interact with the reflecting boundary before exhausting their budgets.
Although additional walks reduce noise predictably and run efficiently in parallel, adapting variance reduction methods from steady-state Monte
Carlo solvers remains an important direction \citep{Qi:2022:Bidirectional,Miller:2023:Boundary,Bakbouk:2023:Mean,Huang:2025:Guiding,Zhou:2025:Harmonic}.
Distributed and GPU implementations could provide further runtime improvements \citep{Sawhney:2026:WoSX}.

Our formulation assumes a linear heat equation with constant diffusivity and mixed Dirichlet--Neumann conditions.
Extending it to variable coefficients and transient Robin or radiative boundary conditions would broaden its scope, building on steady-state methods \citep{Sawhney:2022:Gridfree,Miller:2024:Walkin,Bao:2026:Monte}.
Efficient estimators for spatial, temporal, and parameter derivatives could enable sensitivity analysis, optimization, and inverse problems.
More broadly, we hope these developments help extend grid-free Monte Carlo solvers to other time-dependent PDEs.

\bibliographystyle{ACM-Reference-Format}
\bibliography{strings-full, rendering-bibtex, additional-bibtex}

\end{document}